\documentclass{LMCS}

\def\dOi{11(2:13)2015}
\lmcsheading%
{\dOi}
{1--26}
{}
{}
{May.~19, 2014}
{Jun.~24, 2015}
{}

\ACMCCS{[{\bf Software and its engineering}]: Software notations and
  tools---Formal language definitions---Syntax\,/\,Semantics; [{\bf
    Theory of Computation}]: Logic; Formal languages and automata
  theory---Formalisms---Rewrite systems}

\usepackage[usenames,dvipsnames]{xcolor}
\usepackage[all]{xypic}
\xyoption{pdf}

\usepackage{hyperref,graphicx}
\usepackage{amsmath,amssymb}
\usepackage{color}
\usepackage{vmacros}

\usepackage{tikz}
\usetikzlibrary{arrows,positioning,calc,backgrounds}
\usepackage{pgffor}
\usepackage{gnuplot-lua-tikz}

\usepackage{wrapfig}

\def\MA#1{\begin{bmatrix}#1\end{bmatrix}}
\def\Ga{{\mathit\Gamma}}
\def\Da{{\mathit\Delta}}

\def\hdot{\cdot\hskip-.2ex}

\def\ot{\leftarrow}

\newcommand{\agents}{\mathcal{A}}
\newcommand{\sites}{\mathcal{S}}
\newcommand{\edges}{\mathcal{E}}

\newcommand{\SG}{\mathbf{SG}}

\newcommand{\rSGe}{\mathbf{rSGe}}
\newcommand{\Set}{\mathbf{Set}}

\def\anon#1{|#1|}

\newcommand{\shapes}{\mathcal{P}}
\newcommand{\costs}{\eps}
\newcommand{\generators}{\mathcal{G}}
\newcommand{\LTS}{\mathcal{L}}
\newcommand{\MC}{\mathcal{M}}
\renewcommand{\vec}[1]{\boldsymbol{\mathbf{#1}}}

\newcommand{\cyc}[1]{\hdot #1 \cdot}

\def\mOR{{\mathit\Upsilon}}

\def\PR{\emph{Proof. }}
\def\RP{\qed}

\newcommand{\set}[1]{\left\{ #1 \right\}}

\tikzstyle{rule}=[thick,->]

\newcommand{\birulearrow}[2]{\hspace{1ex}\tikz{%
  \draw[rule] (0, .15) -- node[above] {#1} ++( .5, 0);
  \draw[rule] (.5,  0) -- node[below] {#2} ++(-.5, 0);
}\hspace{1ex}}
\newcommand{\tofrom}{\leftrightharpoons}

\newcommand{\arrsn}[3][]{\arr[#1]{#2.south}{#3.north}}
\newcommand{\arr}[3][]{%
  \draw[-bigto,#1] ($(#2)!.1!(#3)$) -- ($(#2)!.9!(#3)$);}

\newcommand{\n}[3][n]{\node[#1] (#2) at (#3) {}}
\newcommand{\e}[3][e]{\draw[#1] (#2) -- (#3)}

\tikzstyle{grphnode}=[rectangle,grphdiag-bg]
\tikzstyle{ingrphdiag}=[>=stealth,semithick,
  n/.append style={semithick,minimum size=4pt}]
\tikzstyle{relevant}=[fill=green]

\tikzstyle{grphdiag-bg}=[rounded corners, fill=white!85!black]

\tikzstyle{grphdiag}=[>=stealth, framed, thick,%
  background rectangle/.style=grphdiag-bg]

\tikzstyle{n}=[circle,draw=Black]
\tikzstyle{n1}=[n,fill=Orange]
\tikzstyle{n2}=[n,fill=Blue]
\tikzstyle{n3}=[n,fill=White]
\tikzstyle{n4}=[n,fill=White!85!Black]
\tikzstyle{e}=[thick]

\pgfarrowsdeclare{bigto}{bigto}
{
  \arrowsize=-1.54pt%
  \advance\arrowsize by -1.23\pgflinewidth%
  \pgfarrowsleftextend{+\arrowsize}
  \arrowsize=0.28pt%
  \advance\arrowsize by .55\pgflinewidth%
  \pgfarrowsrightextend{\arrowsize}
}
{
  \arrowsize=0.56pt%
  \advance\arrowsize by .5\pgflinewidth%
  \pgfsetlinewidth{0.8\pgflinewidth}
  \pgfsetdash{}{+0pt}
  \pgfsetroundcap
  \pgfsetroundjoin
  \pgfpathmoveto{\pgfqpoint{-3\arrowsize}{4\arrowsize}}
  \pgfpathcurveto
  {\pgfqpoint{-2.75\arrowsize}{2.5\arrowsize}}
  {\pgfqpoint{-1\arrowsize}{0.5\arrowsize}}
  {\pgfqpoint{0.5\arrowsize}{0pt}}
  \pgfpathcurveto
  {\pgfqpoint{-1\arrowsize}{-0.5\arrowsize}}
  {\pgfqpoint{-2.75\arrowsize}{-2.5\arrowsize}}
  {\pgfqpoint{-3\arrowsize}{-4\arrowsize}}
  \pgfusepathqstroke
}

\title[Thermodynamic graph-rewriting]{
Thermodynamic graph-rewriting
}

\author[V.~Danos]{Vincent Danos\rsuper a}
\address{{\lsuper{a,c}}School of Informatics, University of Edinburgh}
\email{vdanos@inf.ed.ac.uk, r.honorato@sms.ed.ac.uk}

\author[R.~Harmer]{Russ Harmer\rsuper b}
\address{{\lsuper b}LIP, CNRS \& ENS Lyon}
\email{russell.harmer@ens-lyon.fr}

\author[R.~Honorato-Zimmer]{Ricardo Honorato-Zimmer\rsuper c}
\address{\vspace{-18 pt}}


\begin{document}

\begin{abstract}
We develop a new {thermodynamic} approach to stochastic graph-rewriting. The ingredients are a finite set of reversible graph-rewriting rules $\generators$
(called generating rules), a finite set of connected 
graphs $\shapes$ (called energy patterns), 
and an energy cost function $\eps$ which associates 
real values to each of these energy patterns. 
The idea is that $\generators$ 
defines the qualitative dynamics, by showing which transformations are possible,
while 
$\shapes$ and $\costs$ allow one to attach an energy to the reachable
graphs and, thereby, describe their long-term probability distribution $\pi$.
Given $\generators$ and $\shapes$, we construct a finite set of rules $\generators_{\shapes}$ 
which (i) has the same qualitative transition system as $\generators$; and (ii)
when equipped with rates according to $\costs$, 
defines a continuous-time Markov chain 
of which $\pi$ is the unique fixed point.
The construction relies on the use of site graphs and a technique of `growth policy' for quantitative rule refinement 
which is of independent interest. 
%

This division of labour between the qualitative and long-term
quantitative aspects of the dynamics leads to intuitive and concise
descriptions for realistic models (see \S\ref{alloring}).
It also guarantees thermodynamical consistency (\emph{aka} detailed
balance), otherwise known to be undecidable, which is important for
some applications. Finally, it leads to parsimonious parameterizations
of models, again an important point in some applications.


\end{abstract}

\maketitle

\section{Introduction}
%

Along with Petri nets, communicating finite state machines, and process algebras, models of concurrent systems based on
graphs and graph transformations (GTS) have long been investigated as means to describe, verify and synthesize distributed systems~\cite{handbook}. Beyond their visual aspect, which is often useful in modelling situations, there is a lot 
to like about GTSs: there are category-theoretic frameworks to express them and encapsulate their syntax; and the existence of a strong meta-theory~\cite{sobo} is a reassurance that methodologies developed in specific cases can be `ported' to other variants. 

Graph-rewriting rules are convenient for writing compact models and modifying them~\cite{agile}, and lend themselves naturally to probabilistic extensions~\cite{heckel,krivinemilner}. However, for all their flexibility, even rules can only do so much. We ask in this paper ``what if we did not have to write the rules?''. This is where we take a page from the book of classical statistical mechanics. In such models, which often involve graph-like structures, as in the Ising model, the dynamics is not described upfront. Instead, the system of interest is equipped with an `energy lansdcape' which  specifies its long run behaviour, be it deterministic as in classical mechanics, or probabilistic in statistical physics. The dynamics just follows from the energy data. In the eye of a computer scientist, this use of energy looks like a latent syntax. (This is especially true in the application of these ideas to molecular dynamics.)

The broad aim of this paper is to make this syntax explicit by
introducing {energy patterns} and {costs} from which the total energy
of a state of the system can be computed; and to define a procedure
whereby, indeed, the dynamics described as 
probabilistic graph-rewriting rules can be derived from these energy
data.  Descriptively, this takes us to an entirely new level of
conciseness (as in the example shown in \S\ref{alloring}).
It also guarantees thermodynamical consistency by construction,
a property known otherwise to be undecidable~\cite{et1}.
This property provides a predictable equilibrium state for the system
which can be used as a guide when writing models by giving an
intuitive understanding of favoured states and the (possibly
overlapping) subgraphs within them. 
But perhaps the nicest byproduct of this approach is the fact that the
methodology leads to parsimonious parameterizations.  The parameter
space which usually scales as the number of rules (which in turn has
at best a logarithmic impact on the cost of a simulation
event~\cite{scalable}), will now scale as the number of energy
patterns provided in the specification.

The particular GTSs we consider form a reversible subset of the Kappa site-graph stochastic rewriting language. 
Kappa is used for the simulation and analysis of combinatorial dynamical systems as typically found in cellular signalling networks%
~\cite{pysb,krantz} and has been predicted to ``become one of the mainstream modelling tools of systems biology within the coming decade''~\cite{naturemethods}. 
Similar graph formalisms where nodes have a controlled valence/degree have been considered 
\eg\ the BNG language~\cite{Faeder,Fontana}, Kissinger and Dixon's quantum proof language~\cite{dixon}, and Kirchner \emph{et al.} chemical calculi~\cite{kirchner}. Site-graph rewriting has found recently a `home' both in the single-pushout GTS tradition~\cite{jon1}
and the double-pushout one~\cite{kappadpo,jonandtobias}. 
This makes one hopeful that the thermodynamic methodology we propose can cross over to other fields where quantitative GTSs can be used, \eg\ in the modelling of adaptive networks~\cite{gross}. While our scalable energy-based parameterization is particularly important in biological applications where parameters often need to be inferred, 
one can imagine it to be useful in other modelling situations with uncertainty.

\paragraph{Outline:}
We start by introducing a running example that will help us illustrate
the concepts presented throughout the paper. Then we proceed with the definition 
and relevant properties of the specific GTS we use, namely a simple reversible 
fragment of Kappa. Next, we introduce growth policies (adapted from
Ref.~\cite{gp2010}), a tool which allows one to replace a rule with an
orthogonal set of refined rules while preserving the quantitative
semantics. We use this tool with a specific policy which refines a
rule into finitely many rules, each of which has a definite energy
balance with respect to a given set of energy patterns. This leads to
our main theorem which guarantees that the stochastic dynamics of the
obtained refined rule set 
converges to an equilibrium distribution parametrized by the cost of
each energy pattern. Throughout, the presentation is set in
category-theoretic terms and mostly self-contained.
A substantial example concludes the paper.

\subsection{Running example: Assembling triangles}
\label{triangles}
The following example will be used to show the intrincacies of
our rule generation mechanism.
We consider 
graphs with three types of nodes 
where each can only bind nodes of a different type,
and at most one thereof.
As generators, we consider simple binding and
unbinding rules subject to this constraint
(see details in \S\ref{rules}):

\medskip
\begin{center}
  \begin{tikzpicture}[grphdiag]
    \n[n1]{a}{0,0};
    \n[n2]{b}{.8,0};
  \end{tikzpicture}
  \birulearrow{}{}
  \begin{tikzpicture}[grphdiag]
    \n[n1]{a}{0,0};
    \n[n2]{b}{.8,0};
    \e{a}{b};
  \end{tikzpicture}
\end{center}
\begin{center}
  \begin{tikzpicture}[grphdiag]
    \n[n2]{b}{0,0};
    \n[n3]{c}{.8,0};
  \end{tikzpicture}
  \birulearrow{}{}
  \begin{tikzpicture}[grphdiag]
    \n[n2]{b}{0,0};
    \n[n3]{c}{.8,0};
    \e{b}{c};
  \end{tikzpicture}
\end{center}
\begin{center}
  \begin{tikzpicture}[grphdiag]
    \n[n3]{c}{0,0};
    \n[n1]{a}{.8,0};
  \end{tikzpicture}
  \birulearrow{}{}
  \begin{tikzpicture}[grphdiag]
    \n[n3]{c}{0,0};
    \n[n1]{a}{.8,0};
    \e{c}{a};
  \end{tikzpicture}
\end{center}

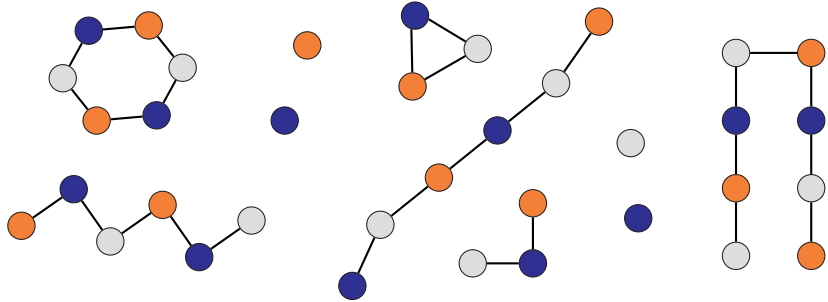
\begin{figure}[b!]
  \bigskip
  \hspace{.7cm}
  \begin{tikzpicture}
    \useasboundingbox (-.2,0) rectangle (10.3,3.5);
    \begin{scope}[transform canvas={shift={(-.2,.8)},rotate=-10}]
      \n[n1]{b1}{0,0};
      \n[n2]{b2}{.6,.6};
      \n[n4]{b3}{1.2,0};
      \n[n1]{b4}{1.8,.6};
      \n[n2]{b5}{2.4,0};
      \n[n4]{b6}{3,.6};
      \e{b1}{b2.south west};
      \e{b2.south east}{b3};
      \e{b3}{b4.south west};
      \e{b4.south east}{b5};
      \e{b5}{b6.south west};
    \end{scope}
    \begin{scope}[transform canvas={shift={(.8,2.2)},rotate=5}]
      \n[n1]{c1}{0,0};
      \n[n2]{c2}{.8,0};
      \n[n4]{c3}{1.2,.6};
      \n[n1]{c4}{.8,1.2};
      \n[n2]{c5}{0,1.2};
      \n[n4]{c6}{-.4,.6};
      \e{c1}{c2};
      \e{c2}{c3};
      \e{c3}{c4};
      \e{c4}{c5};
      \e{c5}{c6};
      \e{c6}{c1};
    \end{scope}
    \begin{scope}[transform canvas={shift={(4.8,3)},rotate=30}]
      \n[n1]{a1}{0,-.4};
      \n[n2]{a2}{.5,.4};
      \n[n4]{a3}{1,-.4};
      \e{a1}{a2.south west};
      \e{a1}{a3};
      \e{a3}{a2.south east};
    \end{scope}
    \begin{scope}[shift={(3.3,2.2)}]
      \n[n1]{e1}{0.3,1};
      \n[n2]{e2}{0,0};
    \end{scope}
    \begin{scope}[transform canvas={shift={(4.2,0)},rotate=2}]
      \n[n2]{f1}{0,0};
      \n[n4]{f2}{.4,.8};
      \n[n1]{f3}{1.2,1.4};
      \n[n2]{f4}{2,2};
      \n[n4]{f5}{2.8,2.6};
      \n[n1]{f6}{3.4,3.4};
      \e{f1}{f2};
      \e{f2}{f3};
      \e{f3}{f4};
      \e{f4}{f5};
      \e{f5}{f6};
    \end{scope}
    \begin{scope}[shift={(5.8,.3)}]
      \n[n4]{d1}{0,0};
      \n[n1]{d2}{.8,.8};
      \n[n2]{d3}{.8,0};
      \e{d1}{d3};
      \e{d3}{d2};
    \end{scope}
    \begin{scope}[shift={(8,.9)}]
      \n[n2]{h1}{0,0};
      \n[n4]{h2}{-0.1,1};
    \end{scope}
    \begin{scope}[shift={(9.3,.4)}]
      \n[n4]{g1}{0,0};
      \n[n1]{g2}{0,.9};
      \n[n2]{g3}{0,1.8};
      \n[n4]{g4}{0,2.7};
      \n[n1]{g5}{1,2.7};
      \n[n2]{g6}{1,1.8};
      \n[n4]{g7}{1,.9};
      \n[n1]{g8}{1,0};
      \e{g1}{g2};
      \e{g2}{g3};
      \e{g3}{g4};
      \e{g4}{g5};
      \e{g5}{g6};
      \e{g6}{g7};
      \e{g7}{g8};
    \end{scope}
  \end{tikzpicture}
  \medskip
  \caption{An example of a mixture of chains and $3n$-gons which our three rules can reach starting with 12 agents of each type.}
  \label{mixture}
\end{figure}

\medskip
These rules, given an unbounded supply of the three types of nodes,
can create chains (of any length) and closed cycles (of length some
multiple of three) as shown in Fig.~\ref{mixture}.  Given the
simplicity of the graphs that can be formed, we can describe them
using a linear textual notation where numbers are used to represent
the three types of nodes, then chains are simply written as words
(\eg\ $2312$) while for cycles we indicate half-edges at both ends,
\eg\ $\cyc{123}$ is the triangle and $\cyc{123123}$ is the hexagon.
We will use that shorthand notation in the sequel.

Our goal here is to be able to favour the formation at equilibrium of
certain structures like the triangle by assigning a significantly
negative energy to them (the convention is that a lower energy 
means a higher probability).

\section{Site graph rewriting}\label{rewriting}

\subsection{Site graphs and homomorphisms}\label{sg}

A \emph{site graph} $G$ consists of finite sets of \emph{agents}
(nodes) and \emph{sites} (connection slots), $\agents_G$ and
$\sites_G$, a partial function $\sig_G: \sites_G \rightharpoonup
\agents_G$, and a symmetric \emph{edge} relation $\edges_G$ on
$\sites_G$.
The pair $\sites_G$, $\edges_G$ form an undirected graph; sites not in the domain of $\edges_G$ are said to be \emph{free}. The role of the additional map $\sig_G$ is to assign sites to agents; sites not in the domain of $\sig_G$ are said to be \emph{dangling}, and will be used to represent half-edges (see below). Usually one also endows agents and/or sites with states (as we do in the example treated in  
\S\ref{alloring}); our main construction 
in \S\ref{rulegen} carries over readily to these.

One says $G$ is \emph{realizable} iff
(i) sites have at most one incident edge;
(ii) no dangling site is free; and,
(iii) edges have at most one dangling site.
%
%
Note that point (i) implies that no site has an edge to itself.
\enlargethispage{\baselineskip}

\begin{wrapfigure}[6]{r}{0.32\textwidth}
\vspace{-5ex}
\begin{minipage}{0.32\textwidth}
\AR{
\xymatrix@R=35pt@C=35pt{
\sites_G
\ar[r]^{h_\sites}
\ar@_{->}[d]_{\sig_G}
\ar@{}[dr]|\leq
&
\sites_{G'}
\ar@_{->}[d]^{\sig_{G'}}
\\
\agents_G
\ar[r]_{h_\agents}
&
\agents_{G'}
}
}
\end{minipage}
\end{wrapfigure}

A \emph{homomorphism} $h : G \to G'$ of site graphs is a pair of functions, $h_\sites : \sites_G \to \sites_{G'}$ and $h_\agents : \agents_G \to \agents_{G'}$, such that (i) whenever $h_\agents(\sig_G(s))$ is defined, so is $\sig_{G'}(h_\sites(s))$ and they are equal; and (ii) if $s \mathbin{\edges_G} s'$ then $h_\sites(s) \mathbin{\edges_{G'}} h_\sites(s')$.

A homomorphism $h : G \to G'$ is an \emph{embedding} iff (i) $h_\agents$ and $h_\sites$ are injective; and
(ii) if $s$ is free in $G$, so is $h_\sites(s)$ in $G'$.
%
If $h : G \to G'$ is an embedding and $G'$ is realizable then $G$ is also realizable.

Site graphs and homomorphisms form a category $\SG$ with the natural `tiered' composition; embeddings form a subcategory; if in addition, we restrict objects to
be realizable, we get the subcategory $\rSGe$ of realizable site graphs and embeddings.

Returning to our chains-and-cycles example, we can use sites as a
means to enforce on nodes our earlier constraint where a node can
connect to at most one node of each other type.  Here we present a
possible encoding of a simple chain of 3 nodes:

\bigskip
\begin{center}
  $G = \; \;$
  \begin{tikzpicture}[grphdiag,baseline=-2]
    \path[use as bounding box] (-0.8,0.4) rectangle (3.8,-0.42);
    \node[n3,inner sep=2pt] (x) at (0,0) {$x$};
    \node[n3,inner sep=2pt] (y) at (1.5,0) {$y$};
    \node[n3,inner sep=2pt] (z) at (3,0) {$z$};
    \draw (x) -- (-0.5,0);
    \node at (-0.33,-0.2) {\small $l_x$};
    \draw (x) -- (y);
    \node at (0.42, 0.20) {\small $r_x$};
    \node at (1.13,-0.25) {\small $l_y$};
    \draw (y) -- (z);
    \node at (1.92, 0.19) {\small $r_y$};
    \node at (2.69,-0.23) {\small $l_z$};
    \draw (z) -- (3.5,0);
    \node at (3.36, 0.19) {\small $r_z$};
  \end{tikzpicture}
\end{center}
\smallskip
\begin{alignat*}{3}
  \agents_G &= \set{x, y, z}, &
  \sig_G &= \set{s_a \mapsto a: s \in \set{l, r}, a \in \agents_G}, \\
  \sites_G &= \set{l_x, r_x, l_y, r_y, l_z, r_z}, \quad &
  \edges_G &= \set{(r_x, l_y), (l_y, r_x), (r_y, l_z), (l_z, r_y)}.
\end{alignat*}

Note that this site graph is realizable and
the $l_x$ and $r_z$ sites are free.
To complete the encoding of our example
we need types for nodes which we now introduce.

\subsection{The category of site graphs over \texorpdfstring{$C$}{C}}\label{types}
A homomorphism $h: G \to C$ is a \emph{contact map} over $C$ iff (i) $G$ is realizable, (ii) $\sig_C$ is total and (iii) whenever $h_\sites(s_1)=h_\sites(s_2)$ and $\sig_G(s_1)=\sig_G(s_2)$, then $s_1=s_2$. The third condition of local injectivity means that every agent of $G$ has at most one copy of each site of its corresponding agent in $C$;
$C$ is called the \emph{contact graph}.


\begin{wrapfigure}[6]{r}{0.28\textwidth}
\vspace{-3ex}
\begin{minipage}{0.28\textwidth}
\AR{
\xymatrix@R=25pt@C=25pt{
G
\ar[dr]_{h}
\ar[rr]^{\psi}
&
&
G'
\ar[dl]^{h'}
\\
&
C
}
}
\end{minipage}
\end{wrapfigure}
 
Hereafter, we work in the (comma) category $\rSGe_C$ whose objects are contact maps over $C$, and arrows are embeddings such that the associated triangle commutes in $\SG$.  We write $\mOR_C(h, h')$ for the set of such embeddings between $h$, $h'$ contact maps over $C$; we also write $\anon{\_}$ for the domain functor from $\rSGe_C$ to $\rSGe$ which forgets types. In particular, if $h: G \to C$ is a contact map, we write 
$\anon h$ for its domain $G$.

Note that in $\rSGe$, an embedding $h: G \to G'$ may map a dangling site
$s$ of $G$ to any site $s'$ of $G'$ (provided $h_\sites$ is injective).
In particular, $s'$ does not have itself to be dangling.
This means that dangling sites can be used as an ``any site'' wild card when matching $G$.
In the typed case, \ie\ in $\rSGe_C$, the contact map $h: G \to C$
tells us which agent in $C$ the dangling site belongs to because
$\sig_C$ is total, and this must be respected by $h$.
Dangling sites can now be used as ``binding types'' to express the property of being bound to the site $s$ of an agent of a given type.

The contact graph $C$ is fixed and plays the role of a \emph{type}:
it specifies the kinds of agents that exist, the sites that they may
possess, and which of the 
$|\sites_C|(|\sites_C|+1)/2$ possible edge types are actually valid.
It also gives canonical names to the types of agents and their sites.


In our running example we use the triangle $C = \cyc{123}$ as contact
graph:

\begin{minipage}{.5\textwidth}
  \begin{align*}
    \agents_C &= \set{1, 2, 3}, \quad
    \sites_C = \set{l_1, r_1, l_2, r_2, l_3, r_3}, \\
    \sig_C &= \set{s_a \mapsto a: s \in \set{l, r}, a \in \agents_C}, \\
    \edges_C &= \set{(r_1, l_2), (l_2, r_1), (r_2, l_3),
      (l_3, r_2), (r_3, l_1), (l_1, r_3)}
  \end{align*}
\end{minipage}
\begin{minipage}{.3\textwidth}
  \vspace{3ex}
  \begin{center}
    \begin{tikzpicture}[grphdiag]
      \n[n1]{n1}{0,-.4};
      \n[n2]{n2}{.5,.4};
      \n[n3]{n3}{1,-.4};
      \e{n1}{n2.south west};
      \e{n1}{n3};
      \e{n3}{n2.south east};
    \end{tikzpicture}
  \end{center}
\end{minipage}
\begin{minipage}{.18\textwidth}
  \hfill
\end{minipage}

\bigskip
This contact graph constrains agents to bind only agents of a different
type.  The local injectivity condition on contact maps restricts
the number of sites available for each node to two.
Our simple chain of three nodes from the previous section (\S\ref{sg})
can thus be typed 
using a contact map
$h: G \to C$ with
$h_\agents = \set{x \mapsto 1, y \mapsto 2, z \mapsto 3}$ and
$h_\sites = \{ s_x \mapsto s_{h_\agents(x)}: s \in \set{l, r}, 
x \in \agents_G \}$.

We now extend the shorthand notation introduced in \S\ref{triangles}
to cover the case when $h_\sites$ is not locally surjective. Specifically, 
when a node
does not mention its two allowed sites,
we write `$?$' for the missing site.
A missing site can be introduced by an embedding
and can be free or bound.
Sites do not explicitly show in this notation but they are implicitly
positioned at the left ($l$) and right ($r$) of each agent.
Binding types are denoted by an exponent.
Thus, for instance, in ${}^312?$ we have two agents $u, v$ of type
$1, 2$, respectively, where $u$ is bound to a dangling site of type
$r_3$ on its $l_u$ site and bound to $v$'s $l_v$ site on its $r_u$
site. Similarly, $v$ is bound to $u$ on its $l_v$ site and does not
have any $r$ site.
Using this extended notation we can write down our three rules in text:
$?1 + 2? {}\tofrom{} ?12?$, $?2 + 3? {}\tofrom{} ?23?$, and
$?3 + 1? {}\tofrom{} ?31?$.

Recall that the goal of this example is to control the type of cycles
which one finds at equilibrium. To this effect, we introduce a set of
energy patterns $\shapes$ consisting of one contact map for each edge
type (\ie~$?12?$, $?23?$, and $?31?$), and one for the triangle
$\cyc{123}$.
As we will see, a careful choice of energy costs for each pattern will
indeed lead to a system producing almost only triangles in the long
run (see Fig.~\ref{triangles-sim} in Appendix~\ref{app:triangles}).

\subsection{Multi-sums in \texorpdfstring{$\rSGe_C$}{rSGeC}}
The category $\SG$ has all pull-backs, constructed from those in $\Set$; it is easy to see that they restrict to $\rSGe_C$. $\SG$ also has all push-outs and all sums, but these do not in general restrict to $\rSGe_C$. (Just as push-outs and sums in $\Set$ do not restrict to the subcategory of injective functions.) 


\begin{wrapfigure}[7]{r}{0.37\textwidth}
\begin{minipage}{0.37\textwidth}
\AR{
\xymatrix@R=35pt@C=35pt{
h_1
\ar[r]^{\ta_1}
\ar[dr]_{\ga_1}
&
s_i
\ar@{.>}[d]_{\exists! m}
&
h_2
\ar[l]_{\ta_2}
\ar[ld]^{\ga_2}
\\
&
h
}
}
\end{minipage}
\end{wrapfigure}

\pagebreak

However, $\rSGe_C$ has \emph{multi-sums} meaning that,
for all pairs of site graphs of type $C$,
$h_1: G_1 \to C$ and $h_2: G_2 \to C$,
there exists a family 
of co-spans $\ta^i_1: h_1 \to s_i \ot h_2 :\ta^i_2$,
such that any co-span $\ga_1: h_1 \to h \ot h_2 :\ga_2$
factors through \emph{exactly one} of the family 
and does so \emph{uniquely}.
In outline, given a co-span $\ga_1: h_1 \to h \ot h_2 :\ga_2$,
we first take its pull-back (which is guaranteed to remain within
$\rSGe_C$) and then the push-out of that:
the fact that the original co-span exists implies that
this push-out also remains within $\rSGe_C$.
The multi-sum is then a choice of push-out
for each (isomorphism class of) pull-back.

The idea is that the pairs $\ta^i_1$, $\ta^i_2$ enumerate all minimal
ways in which one can glue $h_1$ and $h_2$, that is to say all the
minimal gluings of $G_1$ and $G_2$ that respect $C$.
There are \emph{finitely} many,
all of which factor through the standard sum
in the larger slice category $\SG_C$.

The notion of multi-sum dates back to Ref.~\cite{diers};
we will call them \emph{minimal gluings} in $\rSGe_C$
according to their intuition in this concrete context,
and use them in \S\ref{pbg} to construct balanced rules.

To illustrate this idea, let us consider, in the context of our
example, the minimal gluings of $D_1 ={} ?123?$ and $D_2 ={} ?231?$.
Computing them is a matter of computing all pull-backs.
One such pull-back is $D_1 \gets{} ?23? {}\to D_2$, 
which gives us the minimal gluing $?1231?$.
On the other hand,
the span $D_1 \gets{} ?3? {}\to D_2$ is not a pull-back
since for all cospans that can close the square,
the span factors through $D_1 \gets{} ?23? {}\to D_2$.
This is a consequence of $?3?$ not being a maximal overlap
of $D_1$ and $D_2$: whenever $3$ is contained in a possible
overlap of $D_1$ and $D_2$, $2$ will also be there.

All minimal gluings of $D_1$ and $D_2$ are displayed
in the following diagram with their respective pullbacks.
The left square has the empty graph as pullback
and $?123? {}+{} ?231?$ as pushout,
the central square has $?23?$ and $?1231?$,
and the right square has $?23? {}+{} ?1?$ and $\cyc{123}$.
Each square uses arrows of different colour and style.

\bigskip
\begin{center}
  \begin{tikzpicture}[thick]
    \node[grphnode] (d1) at (-4,0) {
      \tikz[ingrphdiag]{
        \n[n1]{n1}{0,0};
        \n[n2]{n2}{1,0};
        \n[n3]{n3}{2,0};
        \draw[e] (n1) -- (n2);
        \draw[e] (n2) -- (n3);
      }};

    \node[grphnode] (d2) at (4,0) {
      \tikz[ingrphdiag]{
        \n[n2]{n2}{0,0};
        \n[n3]{n3}{1,0};
        \n[n1]{n1}{2,0};
        \draw[e] (n2) -- (n3);
        \draw[e] (n3) -- (n1);
      }};

    \node[grphnode] (pb1) at (-2.8,2) {
      \tikz[ingrphdiag]{
        \node {$\varnothing$};
      }};

    \node[grphnode] (pb2) at (-.75,2) {
      \tikz[ingrphdiag]{
        \n[n2]{n2}{0,0};
        \n[n3]{n3}{1,0};
        \draw[e] (n2) -- (n3);
      }};

    \node[grphnode] (pb3) at (2,2) {
      \tikz[ingrphdiag]{
        \n[n2]{n2}{0,0};
        \n[n3]{n3}{1,0};
        \n[n1]{n1}{2,0};
        \draw[e] (n2) -- (n3);
      }};

    \node[grphnode] (po1) at (-3.2,-2) {
      \tikz[ingrphdiag]{
        \n[n1]{n1}{0,.3};
        \n[n2]{n2}{1,.3};
        \n[n3]{n3}{2,.3};
        \draw[e] (n1) -- (n2);
        \draw[e] (n2) -- (n3);
        \n[n2]{n4}{0,-.3};
        \n[n3]{n5}{1,-.3};
        \n[n1]{n6}{2,-.3};
        \draw[e] (n4) -- (n5);
        \draw[e] (n5) -- (n6);
      }};

    \node[grphnode] (po2) at (.15,-2) {
      \tikz[ingrphdiag]{
        \n[n1]{n1}{0,0};
        \n[n2]{n2}{1,0};
        \n[n3]{n3}{2,0};
        \n[n1]{n4}{3,0};
        \draw[e] (n1) -- (n2);
        \draw[e] (n2) -- (n3);
        \draw[e] (n3) -- (n4);
      }};

    \node[grphnode] (po3) at (3,-2) {
      \tikz[ingrphdiag]{
        \n[n1]{n1}{0,-.4};
        \n[n2]{n2}{.5,.4};
        \n[n3]{n3}{1,-.4};
        \draw[e] (n1) -- (n2.south west);
        \draw[e] (n1) -- (n3);
        \draw[e] (n3) -- (n2.south east);
      }};

    \arrsn[OliveGreen,dashed]{pb1}{d1};
    \arrsn[OliveGreen,dashed]{pb1}{d2};
    \arrsn[OliveGreen,dashed]{d1}{po1};
    \arrsn[OliveGreen,dashed]{d2}{po1};
    \arrsn[Brown,densely dotted]{pb2}{d1};
    \arrsn[Brown,densely dotted]{pb2}{d2};
    \arrsn[Brown,densely dotted]{d1}{po2};
    \arrsn[Brown,densely dotted]{d2}{po2};
    \arrsn[Plum,dashdotted]{pb3}{d1};
    \arrsn[Plum,dashdotted]{pb3}{d2};
    \arrsn[Plum,dashdotted]{d1}{po3};
    \arrsn[Plum,dashdotted]{d2}{po3};
  \end{tikzpicture}
\end{center}

\pagebreak

\subsection{Rules}\label{rules}

\begin{wrapfigure}[5]{r}{0.32\textwidth}
\vspace{-1ex}
\begin{minipage}{0.32\textwidth}
\AR{
\xymatrix@R=35pt@C=30pt{
L
\ar[dr]_{r_L}
\ar@{.>}[rr]^{I,I}
&
&
R
\ar[dl]^{r_{R}}
\\
&
C
}
}
\end{minipage}
\end{wrapfigure}

A \emph{rule} $r$ over $C$ is a pair of contact maps $r_L: L \to C$,
$r_R: R \to C$ which differ only in their edge structures,
\ie\ $\agents_L=\agents_R$, $\sites_L=\sites_R$, $\sig_L=\sig_R$,
$r_{L,\agents}=r_{R,\agents}$ and $r_{L,\sites}=r_{R,\sites}$.

For example, the unbinding rule $g_{12}=?12? {}\to{} ?1 + 2?$
may be represented as:
\AR{
\agents_L &=& \agents_R &=& \set{u,v}\\
\sites_L  &=& \sites_R  &=& \set{r_u, l_v} \\
\sig_L    &=& \sig_R    &=& \set{r_u \mapsto u, l_v \mapsto v} \\
r_{L,\agents} &=& r_{R,\agents} &=& \set{u \mapsto 1, v \mapsto 2} \\
r_{L,\sites} &=& r_{R,\sites} &=& \set{r_u \mapsto r_1, l_v \mapsto l_2} \\
}
with only the edge structures 
$\edges_L = \set{(r_u, l_v), (l_v, r_u)}$ and
$\edges_R =\emptyset$ differing.

A contact map $h: G \to C$ is a \emph{mixture} iff $\sig_G$ is total
(no dangling edge) and $h_\sites$ is locally surjective, \ie\ for all
$a \in \agents_G$, $h_\sites(\sig_G^{-1}(a)) = \sig_C^{-1}(h_\agents(a))$.
In words, a mixture is a \emph{fully-specified} site graph with
respect to the type $C$.

\begin{wrapfigure}[6]{r}{0.4\textwidth}
\vspace{-4.5ex}
\begin{minipage}{0.4\textwidth}
\EQ{
\label{rewritestep}
\xymatrix@R=35pt@C=35pt{
r_L
\ar@{.>}[r]
\ar[d]_{\psi}
&
r_{R}
\ar[d]^{\psi\st}
\\
h
\ar@{.>}[r]
&
h\st
}
}
\end{minipage}
\end{wrapfigure}

An embedding $\psi: r_L \to h$ induces a \emph{rewrite} of the mixture $h$ by modifying the edge structure of the image of $\psi$ from that of $r_L$ to that of $r_R$. The result of rewriting is a new mixture $h\st$, where $\anon{h\st}$ has the same agents and sites as $\anon h$, and an embedding $\psi\st:r_R \to h\st$.  

This type of rewriting can be formalized using double push-out rewriting~\cite{kappadpo}, but with the simple rules considered here, there is no need.
The inverse of $r$, defined as $r\st := (r_R,r_L)$ is also a valid rule;
by rewriting $h\st$ with $r\st$ via $\psi\st$, we recover $h$ and $\psi$.




Given a finite set of rules $\generators$ over $C$,
we define a labelled transition system 
$\LTS_\generators$ on mixtures over $C$:
a transition from a mixture $h$ is a rewriting step 
determined and labelled by an \emph{event} $(r,\psi)$,
as in diagram~(\ref{rewritestep}), with
$r$ in $\generators$ and $\psi$ in $\mOR_C(r,h)$. 
We suppose hereafter that $\generators$ is closed under rule inversion,
\ie\ $\generators = \generators\st$.
Hence, every $(r,\psi)$-transition has an inverse $(r\st,\psi\st)$,
and $\LTS_\generators$ is symmetric.

\subsection{CTMC semantics}
It is not difficult to see that for any rule $r$,
$|\mOR_C(r_L,h)| \leq |\agents_{\anon h}|^{d(r)}$ where $d(r)$ is the
number of connected components in $r_L$.
Hence, $\LTS_\generators$ has finite out-degree, bounded by
$|\generators|\cdot|\agents_{\anon h}|^d$ for some $d$.
Also, as agents are preserved by rules, the (strongly) connected
components of $\LTS_\generators$ are finite.
Given a \emph{rate map} $k$ from $\generators$ to $\mbb R_+$, we can
therefore equip $\LTS_\generators$ with the structure of a
time-homogeneous irreducible continuous-time Markov chain (CTMC),
simply by assigning rate $k(r)$ to an event of the form $(r,\psi)$.
We write $\LTS^k_{\generators}$ for the CTMC thus obtained.



A finite time-homogeneous CTMC $\MC$ has \emph{detailed balance}
for a probability distribution $\pi$ on $\MC$'s state space iff,
for all states $x$ and $y$, $\pi(x) \cdot q(x,y) = \pi(y) \cdot q(y,x)$
where $q(x,y)$ is $\MC$'s transition rate from $x$ to $y$.
This implies, assuming $\MC$ is irreducible, that $\pi$ is the unique
fixed point of the action of $\MC$ to which the probabilistic state of
$\MC$ converges, regardless of the initial state.

%
%
%


In our case, the probability $\pi(x)$ will be proportional to
$\mathrm{e}^{-E(x)}$, where $E(x)$ is the energy of the system as
defined by a set of patterns and their associated costs.


\subsection{Extensions and rule refinement}

We say an embedding $\phi: s \to s'$ is a \emph{prefix} of
\begin{wrapfigure}[5]{r}{0.25\textwidth}
\vspace{-2.5ex}
\begin{minipage}{0.25\textwidth}
\AR{
\xymatrix@R=20pt@C=20pt{
&
s
\ar@{>}[rd]^{\phi}
\ar@{>}[ld]_{\phi'}
\\
s''
&&
s'
\ar@{>}[ll]^{\ta}
}
}
\end{minipage}
\end{wrapfigure}
$\phi': s \to s''$ if there is some embedding $\theta: s' \to s''$
such that $\ta \phi = \phi'$ and write $\psi \leq \phi$ for this.
We refer to a prefix of an epi $\phi : s \to s'$ as an
\emph{extension} of $s$.
In the category of extensions of $s$, a morphism between objects
$\phi : s \to s'$ and $\phi' : s \to s''$ is an embedding
$\ta : s' \to s''$ such that the triangle commutes.
If $\ta$ is an iso we write $\phi \cong_s \phi'$.

Epis of $\rSGe_C$ are characterized as follows~\cite{gp2010}:
suppose $s: G \to C$ and $s': G' \to C$ are contact maps then
$\phi: s \to s'$ of $\rSGe_C$ is an epi iff every
connected component of $G'$ contains at least one agent in the image
of $\phi_\agents$.
Rule application preserves epis and in fact also prefixes of epis:
\begin{lem}
\label{lem:epis}
Let $r = (r_L, r_R)$ be a rule and $\phi: r_L \to x$ be an embedding
with $r_L$, $r_R$, and $x$ contact maps in $\rSGe_C$.
The embedding $\phi\st: r_R \to x\st$ that results from applying
$r$ to $\phi$ is a prefix of an epi iff $\phi$ is.
\end{lem}
\PR
This amounts to proving that $\psi\st \geq \phi\st$ is an epi
if $\psi \geq \phi$ is. 
For this it is enough to consider the case where
the rule adds or deletes exactly one edge.
Rules that modify more than one edge at a time can be decomposed
as sequences of deletions and insertions of edges.
Given that each deletion and insertion preserves the property,
the sequence will preserve it as well.
%
The case of adding an edge is easy,
as the image of $\psi\st$ has fewer connected components to ``touch''.
The case where $r$ deletes an edge can introduce new connected
components, however in this case both ends $u$, $v$ of the deleted
edge must be in $r_L$, so whether the deletion disconnects or not
the codomain of $\psi$, the components of $\psi\st(u)$ and
$\psi\st(v)$ will have a pre-image, namely $u$ and $v$.
\RP

It follows that the category of extensions of $r_L$ and $r_R$
are isomorphic.
Indeed, because we have chosen to restrict to reversible rules,
the full categories under $r_L$ and $r_R$ are evidently isomorphic,
and by the lemma above this correspondence preserves being
prefix-of-an-epi;
if $\phi$ is an extension of $r_L$,
we will write $\phi\st$ for the corresponding extension of $r_R$.



A family of epis $\phi_i: s \to t_i$ 
\emph{uniquely decomposes} $s$, or 
is a \emph{refinement} of $s$,
if, for all mixtures $h$ and embeddings $\psi: s \to h$,
there exists a unique $i$ and $\psi'$ such that $\psi = \psi' \phi_i$. 
This is the basic requirement for a reasonable notion of rule refinement:
it guarantees that the LHS $s$ of a given rule splits into a
non-overlapping and exhaustive collection of more specific cases $t_i$.

In the next section, we will be constructing specific such decompositions in order to produce families of subrules that are compatible with energy patterns,
%
\ie\ each subrule should produce and consume the same number of energy
patterns in each application, regardless of the particular embedding
$\psi \in \mOR_C(r_L,h)$ that the subrule is applied to.


To find such unique decompositions we first recall the growth policy
method \cite{gp2010}, which works by detailing which agents and sites
should be added to $s$.
%
%
Specifically, a \emph{growth policy} $\Ga$ for $s$ is a family of
functions $\Ga_\phi$, indexed by all extensions $\phi: s \to t$,
where $\Ga_\phi$ maps $u \in \agents_{\anon{t}}$ to a subset $\Ga_\phi(u)$ of $\sig_C^{-1}(t_\agents(u))$, \ie\ 
each agent in $\anon{t}$ 
is allocated a subset of the set of sites 
belonging to the agent $t_\agents(u)$ it is mapped to in $C$. 
An agent in $\anon{t}$ may cover some, or all, of these sites or even
completely extraneous sites: if the former, \ie\ for all $u$ in
$\agents_{\anon{t}}$, $t_\sites(\sig_{\anon t}\mo(u))\subseteq \Ga_\phi(u)$,
we say that $\phi$ is \emph{immature}; if for all $u$s, the inclusion
is an equality and $\phi$ is also an epi,
we say that $\phi$ is \emph{mature}; otherwise $\phi$ is said to be \emph{overgrown}. The functions $\Ga_\phi$ must satisfy, for all extensions $\phi$ and $\phi'\geq\phi$, 
the \emph{faithfulness} property, $\Ga_\phi = \Ga_{\phi'} \circ
\psi_\agents$ with $\psi$ such that $\psi \phi = \phi'$;
so a site requested by $\phi$ must be requested by any further
extension.  Additionally, this property forces $\Ga$ to eagerly ask
for all sites that will be eventually requested at any given agent in
the codomain of $\phi$.  If $\phi$ is not overgrown then no $\phi'
\leq \phi$ is overgrown either.  Also, note that the \emph{union} of
two growth policies is itself a growth policy.

Given an $s$ and a growth policy $\Ga$ for $s$, we define $\Ga(s)$ by
choosing one representative per $\cong_s$-isomorphism class of the
set of all extensions of $s$ which are mature according to $\Ga$.

%

The following theorem 
guarantees that factorizations through $\Ga(s)$ are unique when they
exist, but \emph{not} that they necessarily do exist. In the next
section, we will construct a specific growth policy of interest for
which this property of exhaustivity of the decomposition
can be proved by hand. As such, it fulfils our desired criteria of
providing an exhaustive collection of mutually exclusive sub-cases.

\begin{thm}
\label{unidec}
Let $s$ and $x$ be contact maps and $\Ga$ a growth policy for $s$.
If an embedding $\psi: s \to x$ in $\mOR_C(s, x)$ can be decomposed
in two ways as $\ga_1 \phi_1$ and $\ga_2 \phi_2$ with
$\phi_i: s \to t_i$ in $\Ga(s)$ and $\gamma_i: t_i \to x$ then
$\phi_1 = \phi_2$ and $\ga_1 = \ga_2$.
\end{thm}

\begin{wrapfigure}[10]{r}{0.5\textwidth}
\vspace{-4.2ex}
\begin{minipage}{0.5\textwidth}
\EQ{  
\label{gpproof}
\xymatrix@C=20pt@R=20pt{
s
\ar[rrr]^{\phi_1}
\ar[ddd]_{\phi_2}
\ar[dr]|\phi
&&&
t_1\ar[ddd]^{\ga_1}
\ar[ddl]|{\ta_1}
\\
&
p
\ar[urr]|{\pi_1}
\ar[ldd]|{\pi_2}
\\
&&
m
\ar[rd]
\\
t_2\ar[rrr]_{\ga_2}
\ar[rru]|{\ta_2}
&&&
x
}}
\end{minipage}
\end{wrapfigure}

\PR 
Suppose that $\ga_1\phi_1=\ga_2\phi_2$, where $\phi_1$ and $\phi_2$
are mature extensions of $s$ according to $\Ga$ and $\phi_1 \ne \phi_2$.
As shown in diagram~(\ref{gpproof}), we have an inner square formed
by the pull-back $\pi_1$, $\pi_2$, and the minimal gluing $\ta_1$,
$\ta_2$ of $\ga_1$ and $\ga_2$.  Also $\ta_1$ and $\ta_2$ are epis, as
every connected component of $m$ has a pre-image in $t_1$ or $t_2$ and
so also in $s$, since the $\phi_i$s are epis, and so also in the other
of $t_2$ and $t_1$.

If $\ta_1$, $\ta_2$ are not both isomorphisms then there must be a pair
$u$, $z$, consisting of a node in $m$ with pre-images $u_1$, $u_2$ in
$t_1$, $t_2$ and a site $z$ of $u$, such that $z$ has no pre-image in
exactly one of $\ta_1$, $\ta_2$.  Say it is $\ta_2$.  Since $\phi_1$
is not overgrown, $z\in\Ga_{\phi_1}(u_1)$ and, by faithfulness,
$z\in\Ga_\phi((u_1,u_2))$, where $(u_1,u_2)$ is the pull-back
pre-image of $u_1$ and $u_2$.  So again, by faithfulness,
$z\in\Ga_{\phi_2}(u_2)$ which contradicts our original assumption.
So $\ta_1$ and $\ta_2$ are isos.
It follows that $\phi_1 = \phi_2$ as there is
only one representative per $\cong_s$-isomorphism class in $\Ga(s)$.
Finally, $\ga_1 = \ga_2$ because $\phi_1$ is an epi.
\RP


Given a rule $r$ and an extension $\phi: r_L \to t$ of $r_L$,
we write $r_\phi$ for the refined rule associated to $\phi$;
that is, $r_\phi$ is the pair $(t, t\st)$ with
$t\st$ the codomain of the extension $\phi\st$
corresponding to $\phi$.
Given $\Ga$ a growth policy for $r_L$, we write $\Ga(r)$ for the 
family of rules obtained by refining $r$ according to $\Ga$;
that is $\Ga(r)$
is the family of rules $r_\phi$, for $\phi$ ranging in $\Ga(r_L)$.

An example of growth policy is the \emph{ground} policy which assigns all possible sites to all agents. In this case, $\Ga(s)$ is simply the set, possibly infinite, of all epis of $s$ into mixtures, considered up to $\cong_{s}$; and $\Ga(r)$, the {ground refinement} of $r$, contains all refinements of $r$ along those epis, which therefore directly manipulate mixtures.
It is easy to see that the ground refinement for our running example
is infinite, since each of the three rules can trigger the extension
of a chain of any length.

%

\section{Rule generation}\label{rulegen}
We fix a finite set $\generators$ of \emph{generator} rules;
and a finite set $\shapes$ of connected contact maps in $\rSGe_C$;
these are our \emph{energy patterns}.

A canonical set of generator rules can be derived from the contact
graph $C$ by constructing a pair of binding/unbinding rules (and a full set
of state changes if we were to use site graphs with states as
we do in applications) for
each edge in $\edges_C$.  This set is maximally general in that
each generator rule asks for the least possible context to trigger
a binding or unbinding event. This is the set which we have chosen to use in our running example. In the next example which we present
in \S\ref{alloring}, we will consider a strict subset of this 
canonical generating set. But, in general, there is no prescription regarding
which rules one decides to incorporate to $\generators$.

The goal is now to refine $\generators$ into a new rule set $\generators_\shapes$ 
where each refined rule is $\shapes$-\emph{balanced}, which means that, however applied, it 
consumes or produces a fixed amount of each $c$ in $\shapes$. 
The construction proceeds in two steps. Firstly, we characterize balanced refinements; 
secondly, we define a specific growth policy with balanced mature extensions. Using 
Th.~\ref{unidec}, we show that these mature extensions obtain a proper finite refinement of $\generators$.

Note that ground extensions of 
$g$ are trivially balanced but, in general, the ground refinement is impractically large or even infinite; ours will always be finite.




\subsection{Consumption and production of patterns}

Consider $c$ in $\shapes$ and a rule $r$.
%
%
For an $r$-event $\psi$ to \emph{consume} an instance
$\ga$ of $c$ in a mixture $h$, $\ga_\sites$, and $\psi_\sites$
must have images which intersect on at least one site
which is modified by $r$ (by adding an edge if it was free, or
\begin{wrapfigure}[7]{r}{0.32\textwidth}
\vspace{-2ex}
\begin{minipage}{0.32\textwidth}
\AR{
\xymatrix@R=35pt@C=35pt{
c
\ar[r]^{\ga'}
\ar[dr]_{\ga}
&
m
\ar@{.>}[d]
&
r_L
\ar[l]_{\psi'}
\ar[ld]^{\psi}
\\
&
h
}
}
\end{minipage}
\end{wrapfigure}
by deleting or changing an edge it was part of).
This is the case iff the associated minimal gluing
$(\ga',\psi')$ (obtained by restricting the co-span to the union of its
images in $h$) has the same property.
Likewise, for an $r$-event to \emph{produce} an instance of $c$,
the associated minimal gluing between $c$ and $r_R$
must have a modified intersection.
We call such minimal gluings \emph{relevant}; they are the ones which
underlie events that can affect the instances of $c$.

This notion can be illustrated by looking at the minimal gluings
of the left-hand side of the unbinding rule $?12? {}\to{} ?1 + 2?$ and
energy pattern $?1231?$ (supposing for a moment that we had $?1231?$
in $\shapes$): 

\begin{center}
  \begin{tikzpicture}[thick,x=.75cm]
    \node[grphnode] (d1) at (-4,0) {
      \tikz[ingrphdiag]{
        \n[n1]{n1}{0,0};
        \n[n2]{n2}{1,0};
        \n[n3]{n3}{2,0};
        \n[n1]{n4}{3,0};
        \draw[e] (n1) -- (n2);
        \draw[e] (n2) -- (n3);
        \draw[e] (n3) -- (n4);
      }};

    \node[grphnode] (d2) at (4.5,0) {
      \tikz[ingrphdiag]{
        \n[n1]{n1}{0,0};
        \n[n2]{n2}{1,0};
        \draw[e] (n1) -- (n2);
      }};

    \node[grphnode] (pb1) at (-3,2) {
      \tikz[ingrphdiag]{
        \node {$\varnothing$};
      }};

    \node[grphnode] (pb2) at (0,2) {
      \tikz[ingrphdiag]{
        \n[n1]{n1}{0,0};
        \n[n2]{n2}{1,0};
        \draw[e] (n1) -- (n2);
      }};

    \node[grphnode] (pb3) at (3,2) {
      \tikz[ingrphdiag]{
        \n[n1]{n1}{0,0};
      }};

    \node[grphnode] (po1) at (-4,-2) {
      \tikz[ingrphdiag]{
        \n[n1]{n1}{0,.3};
        \n[n2]{n2}{1,.3};
        \n[n3]{n3}{2,.3};
        \n[n1]{n4}{3,.3};
        \draw[e] (n1) -- (n2);
        \draw[e] (n2) -- (n3);
        \draw[e] (n3) -- (n4);
        \n[n1]{n5}{1,-.3};
        \n[n2]{n6}{2,-.3};
        \draw[e] (n5) -- (n6);
      }};

    \node[grphnode,relevant] (po2) at (0,-2) {
      \tikz[ingrphdiag]{
        \n[n1]{n1}{0,0};
        \n[n2]{n2}{1,0};
        \n[n3]{n3}{2,0};
        \n[n1]{n4}{3,0};
        \draw[e] (n1) -- (n2);
        \draw[e] (n2) -- (n3);
        \draw[e] (n3) -- (n4);
      }};

    \node[grphnode] (po3) at (4.5,-2) {
      \tikz[ingrphdiag]{
        \n[n1]{n1}{0,0};
        \n[n2]{n2}{1,0};
        \n[n3]{n3}{2,0};
        \n[n1]{n4}{3,0};
        \n[n2]{n5}{4,0};
        \draw[e] (n1) -- (n2);
        \draw[e] (n2) -- (n3);
        \draw[e] (n3) -- (n4);
        \draw[e] (n4) -- (n5);
      }};

    \arrsn[OliveGreen,dashed]{pb1}{d1};
    \arrsn[OliveGreen,dashed]{pb1}{d2};
    \arrsn[OliveGreen,dashed]{d1}{po1};
    \arrsn[OliveGreen,dashed]{d2}{po1};
    \arrsn[Brown,densely dotted]{pb2}{d1};
    \arrsn[Brown,densely dotted]{pb2}{d2};
    \arrsn[Brown,densely dotted]{d1}{po2};
    \arrsn[Brown,densely dotted]{d2}{po2};
    \arrsn[Plum,dashdotted]{pb3}{d1};
    \arrsn[Plum,dashdotted]{pb3}{d2};
    \arrsn[Plum,dashdotted]{d1}{po3};
    \arrsn[Plum,dashdotted]{d2}{po3};
  \end{tikzpicture}
\end{center}

Of these, only the central square is a relevant minimal gluing since
the image of the edge consumed by the rule is contained in the image
of the energy pattern in the minimal gluing.


\subsection{\texorpdfstring{$\shapes$}{shapes}-balanced extensions}

Given $g$ in $\generators$ with LHS $g_L$ and
$\phi: g_L \to t$ an extension of $g_L$,
we say that $\phi$ is $\shapes$-\emph{left-balanced} iff,
for all relevant minimal gluings $\ga : c \to m \gets t : \ta$
with $c \in \shapes$, $\ta$ is an isomorphism.
This means that the image of $c$ under $\ga$ is contained in $t$.
\begin{wrapfigure}[8]{r}{0.36\textwidth}
\vspace{-3ex}
\begin{minipage}{0.36\textwidth}
\EQ{
\label{gluex}
\xymatrix@R=25pt{
&&
g_L\ar[d]^\phi\\
c\ar[rd]_\ga
&&
t\ar[dl]^{\ta}_{\simeq}\\
&
m
}
}
\end{minipage}
\end{wrapfigure}
Symmetrically, one says that $\phi$ is $\shapes$-\emph{right-balanced}
iff $\phi\st$ is a $\shapes$-left-balanced extension of $r\st$.
An extension $\phi$ is $\shapes$-\emph{balanced}
iff it is $\shapes$-left- and $\shapes$-right-balanced.
We say that a balanced extension $\phi$ is \emph{prime} iff 
it is minimally so, \ie\ any prefix of $\phi$ that is
$\shapes$-balanced 
is isomorphic to $\phi$ (as an extension). Prime extensions are epis since erasing an `untouched' connected component in the codomain preserves balance.


If $\phi$ is a $\shapes$-balanced extension of $g$,
the refined rule $g_\phi$ has a \emph{balance vector} in $\mbb Z^\shapes$, written $\Da\phi$, 
where, for each $c \in \shapes$, $\Da\phi(c)$ is the number of copies of $c$ produced by \emph{any} $g_\phi$-event, which is also the difference between the number of embeddings of $c$ in the RHS and the LHS of $g_\phi$.  
In other words, for any mixture $h$, balance of $\phi$ guarantees $\Da\phi(c)=|\mOR_C(c,h\st)|-|\mOR_C(c,h)|=|\mOR_C(c,g_{\phi,R})|-|\mOR_C(c,g_{\phi,L})|$.
Conversely, if $c$ in $\shapes$ violates the condition of
diagram~(\ref{gluex}), 
well-chosen applications of $g_\phi$
will result in different and context-dependent $\Da\phi(c)$.  Thus, the notion of balanced
extension characterizes the property that we want.

To illustrate these definitions consider the unbinding rule
$g_{12} = ?12? {}\to{} ?1 + 2?$ with the trivial extension 
$\phi = \vec{1}_{r_L}$ (the identity arrow
of the left-hand side of rule $g_{12}$).
Define $\shapes_1 = \set{?12?, ?23?, ?31?}$ and
$\shapes_2 = \set{\cyc{123}}$. On the one hand, $\phi$ is $\shapes_1$-balanced. Of all possible gluings of
patterns in $\shapes_1$ with $\phi$'s codomain, only the 
trivial one of $?12?$ with itself is relevant to $g_{12}$.
On the other hand, $\phi$ is not $\shapes_2$-balanced, because there is 
a gluing between $\cyc{123}$ and $?12?$ 
mapping the $12$ edge to itself in the triangle.
This gluing is relevant since applying $g_{12}$ breaks the triangle;
and clearly $\theta:{} ?12? {}\to \cyc{123}$ is not an iso.  
Failure to be balanced reflects in the fact that different applications 
of $g_{12}$ incur different values of $\Da\phi(\cyc{123})$:
namely $1$ or $0$ depending on whether the broken edge belongs 
to a triangle.

The ideas introduced now are discussed in a detailed example in \S\ref{triangles-detail}.

\subsection{Absorb or Avoid}\label{pbg}
We have just seen how the property of being $\shapes$-balanced characterizes, among all extensions $\phi$ of $g_L$, those with an unambiguous energy balance. The next step is to define a growth policy with a set of mature extensions which are balanced and form a unique decomposition of $g_L$. If we return to diagram~(\ref{gluex}) in the case where $\ta$ is not an isomorphism, the natural idea is to request the addition to the codomain of $\phi$ of these sites which are in $m$ and belong to nodes in the image of  $\theta_\agents$. In this way, we force $\phi$ to grow further and make a decision about whether it wants to absorb the image of $c$ in $m$, or would rather avoid this by growing in a way that is incompatible with the $\ga$, $\ta$ gluing.

Some care is needed to ensure faithfulness,
\ie\ $\Ga_\phi=\Ga_{\phi'\phi}\,\phi'_\agents$,
since relevant minimal gluings on $\phi$ can disappear
along a further extension $\phi'$
so that a site that was requested at $\phi$ may no longer be so
after at $\phi'\phi$. To 
address this, we add site requests from all relevant minimal gluings
in the \emph{past} of an extension.

\begin{wrapfigure}[6]{r}{0.5\textwidth}
\vspace{-5ex}
\begin{minipage}{0.5\textwidth}
\EQ{
\label{gpdiag1}
\xymatrix@R=15pt{
&&
g_L\ar[d]_{\phi_1}
\ar@/^/[ddr]^{\phi}
\\
c\ar[rd]_{\ga}
&&
t_1\ni u_1
\ar[ld]^{\ta}
\ar[rd]_{\phi_2}
\\
&m'
&&
t\ni u
}}
\end{minipage}
\end{wrapfigure}

Given $g$ in $\generators$ we define a growth policy $\Ga$ for $g_L$.
Suppose $\phi: g_L \to t$ is an extension of $g_L$.
We set $\Ga_\phi$ to request a site $s$ in $\sig_C\mo(t_\agents(u))$
at agent $u$ in $\agents_{\anon t}$ iff either
(i) 
$u = \phi_\agents(u_0)$ and $s = \phi_\sites(s_0)$
for some $u_0$ in $\agents_{\anon{g_L}}$, 
$s_0$ in $\sites_{\anon{g_L}}$; or 
(ii) 
$\phi$ factorizes as $\phi_2\phi_1$, where $\phi_1: g_L \to t_1$, 
and there is a relevant minimal gluing $\ga: c \to m' \gets t_1 :\ta$,
with $c$ in $\shapes$, and
some $u_1$ in $\agents_{\anon{t_1}}$
and a site $s'_1$ in $\sig_{\anon{m'}}\mo(\ta_\agents(u_1))$
such that $u = \phi_{2,\agents}(u_1)$ and $s = m'_\sites(s'_1)$.

We refer to the extension $\phi_2: t_1 \to t$ as a \emph{rewind} of
$\phi$; we say that the request of $s$ at $u$ originates from $u_1$.  
The first clause simply ensures that all sites already
covered in $g_L$ are asked for; the second one adds in sites which
appear by gluing at some point between $g_L$ and $t$,
and implements the absorb-or-avoid constraint 
explained beforehand. 

Symmetrically, we define a growth policy $\Ga\st$ for $g_R$ by applying the same definition to the reverse generator $g\st$. Since 
extensions of $g_L$ and $g_R$ are isomorphic, we can, with a slight abuse of notation, 
define $\Ga^\shapes := \Ga \cup \Ga\st$.

\begin{thm}
\label{main}
The above $\Ga^\shapes$ is indeed a growth policy for $g_L$; 
the induced refined family of rules $\Ga^\shapes(g)$ is
exhaustive,
non-empty, 
$\shapes$-balanced,
and finite.
\end{thm}
\PR
We take the same notations as in diagram (\ref{gpdiag1}).

\emph{Growth policy}: Clearly, $\Ga^\shapes_{\phi_1}(u_1)\subseteq\Ga^\shapes_\phi(u)$ as every request for a site in $t_1$
will propagate to $t$ by definition. 
To prove the other direction, we need to verify that the requests generated by rewinds do not depend on the choice of factorization. So, wlog, consider gluings on left extensions of $g$, 
and let an alternative factorization of $\phi$ through $t_2$ be given which 
gives rise to a site request in $u$ originating from some $u_2$ in $t_2$:
\AR{
\xymatrix@R=15pt{
&
g_L
\ar@/_10pt/[ldd]_{\phi_1}
\ar@/^10pt/[rdd]^{}
\ar@{.>}[d]
\\
&
p\ni(u_1,u_2)
\ar[ld]\ar[rd]
\\
t_1\ni u_1
\ar[rd]_{\phi_2}
&&
t_2\ni u_2
\ar[ld]
\\
&t\ni u}
}

Consider $p$ the pull-back of the two rewinds (ie the lower co-span); 
by construction it contains a pre-image for $u_1$ and $u_2$, say $(u_1,u_2)$.
The relevant minimal gluing of $c$ and $t_2$ that makes the site request restricts to another (minimal) gluing of $c$ and $p$. This new gluing is still obviously relevant (as it contains the same overlap with the original $g_L$); as such, the same site request is made by the pre-image $(u_1,u_2)$ agent in $p$ which then propagates to $u_1$ in $t_1$ as required.



\emph{Surjectivity}: Let $\phi: g_L \to x$ be an embedding of $g_L$ into a mixture.  We can restrict the co-domain of $\phi$ to be the connected closure $y$ of the image of $\phi$ in $x$, resulting in an epi $\phi_y : g_L \to y$.
Let us further restrict $y$ by removing: 
1) all unneeded agents, \ie\
those that have no sites requested by the growth policy, 
2) all sites not
requested by the growth policy \emph{except} those bound to sites that
are requested and, finally, 
3) all connected components that no longer
have a pre-image in $g_L$; call the result $z$.  The non-requested sites left
in $z$ act as binding types (see \S\ref{types}).  We thus obtain an epi $\phi_z : g_L
\to z$ which is mature with respect to $\Ga^\shapes$ since, by
construction, it has all the sites requested by $\phi_y$ and so, by
faithfulness, all those requested by $\phi_z$.


\emph{Non-empty}: Clause (i) guarantees that we request at least the sites in $g$ which implies that $g$ is not overgrown. 

\emph{Balanced}: If $\phi \in \Ga^\shapes(g)$ is not balanced then there must be some relevant minimal gluing inducing a further site request, hence $\phi$ cannot be  
mature.

\emph{Finite}:
A request for a site $a$ at some node in an extension 
$\phi:g_L\to t$, or $\phi\st:g_R\to t$, originates from a relevant minimal gluing of some $c$ in $\shapes$ with a prefix $\phi_1$ of $\phi$. Because this gluing is relevant, it must be that $a$ is at a distance from the image of 
$g_L$ in the codomain of $\phi_1$ which is at most $\da(c)$, the diameter of $c$ (else $c$ would not intersect the image of $g_L$). The same bound holds in the codomain of 
$\phi$, as distances can only contract by  further extension. Therefore any site requested in $t$ has a distance to the image 
$\phi(g_L)$ which is bounded by $\max_{c\in\shapes}\da(c)$. If $\phi$ is not overgrown, this sets a bound on the diameter of $t$. Hence there are finitely many mature extensions.
\RP


Therefore, given $\generators$ and $\shapes$, we obtain 
a finite $\shapes$-balanced rule set which refines $\generators$ exhaustively, 
by setting $\generators_{\shapes}:=\dot\cup_{g\in\generators}\Ga^\shapes(g)$ 
(disjoint sum). To every refinement $g_\phi$, corresponds an inverse
refinement $g\st_{\phi\st}$; hence, $\generators_{\shapes}=\generators_{\shapes}\st$
is closed under inversion like $\generators$.

\subsection{Other options considered}
As said earlier, there are other ways to generate a balanced rule set. Beyond the ground extensions, one idea is to use \emph{primes}, \ie\ minimal balanced extensions. Clearly, prime extensions factorize all balanced ones -including the ones we generate by using the absorb-or-avoid growth policy presented above. The problem is that this factorization is not unique: prime extensions do not define in general a valid refinement as distinct sub-cases may overlap.




Another idea is to obtain balanced rules by gluing $\shapes$ directly onto a generator rule $g$, in all possible maximal combinations, rather than on all extensions of $g$ as in the above growth policy. This always reveals enough of the context in which $g$ is applied so as to get $\shapes$-balanced refinements because, by definition, no additional form can be glued further. The problem with this approach is the opposite to that with primes: it does not in general build \emph{enough} rules; so the refinement is valid but not exhaustive.



\subsection{Triangles in detail}
\label{triangles-detail}
We now discuss the concepts introduced in this section 
using the example of the unbinding rule $g_{12} ={} ?12? {}\to{} ?1 + 2?$.
Of the four patterns in $\shapes$, we need only
consider the triangle $T = \cyc{123}$ as the others all trivially glue
or trivially don't.

\begin{wrapfigure}[7]{r}{0.43\textwidth}
\vspace{-1.2ex}
\begin{minipage}{0.43\textwidth}
\AR{
\xymatrix@R=15pt{
&&
?12?\ar[d]
\ar@/^/[ddr]^{\phi_{1,?}}
\\
T\ar[rd]_{\ga}
&&
?12?
\ar[ld]^{\ta}
\ar[rd]_{\phi_{1,?}}
\\
&T
&&
12?
}}
\end{minipage}
\end{wrapfigure}

We apply our growth policy to the extension $\phi_{1,?}:{} ?12? {}\to
12?$ whereby we reveal a new free site in $1$.  
The relevant minimal gluing $T \to T \gets{} ?12?$ disappears
in this extension, as the new site is bound in $T$.  
However, by rewinding back
to $?12?$, our growth policy requires us to reveal
additionally the second site of $2$; as such, $\phi_{1,?}$ is not
mature despite being balanced (indeed minimally so).

Intuitively, the growth policy forces us to reveal additional sites of
$?12?$ because an instance of $g_{12}$ may or may not consume a
triangle and so has an ambiguous energy balance.  The definition of
the growth policy makes it explicit that we have to reveal both of the
hidden sites of $?12?$ and so there are four sub-cases to consider:
$12$, ${}^312$, $12^3$ and ${}^312^3$ (using the notation introduced earlier
for binding types).  The first three cases all
{avoid} the triangle, thereby having definite energy balance,
and are moreover mature with respect to the growth policy; however,
the case of  ${}^312^3$ requires further analysis as it covers both the case of
a triangle and of any chain of length at least $4$.

\begin{wrapfigure}[7]{r}{0.45\textwidth}
\vspace{-3ex}
\begin{minipage}{0.45\textwidth}
\AR{
\xymatrix@R=15pt{
&&
?12?\ar[d]_{\phi_{?3,?}}
\ar@/^/[ddr]
\\
T\ar[rd]_{\ga}
&&
?312?
\ar[ld]^{\ta}
\ar[rd]
\\
&T
&&
?3123?
}}
\end{minipage}
\end{wrapfigure}

Consider then the case of the extension to $?3123?$ where the two
binding types of ${}^312^3$ have been embodied in two \emph{distinct}
$3$ agents.  Any rewind of interest has to go back far enough to glue
$T$; there are two maximal such extensions:
$\phi_{?3,?}:{} ?12? {}\to{} ?312?$ and
$\phi_{?,3?}:{} ?12? {}\to{} ?123?$.
The former requests the hidden site of the left-hand $3$ agent of
$?3123?$; the latter (diagram not shown) requests the hidden site of
the right-hand $3$ agent.
As such, $?3123?$ is not mature with respect to the growth policy and
we must expand it into its four sub-cases:
$3123,{}^23123,3123^1,{}^23123^1$.

Our final refinement, according to the growth policy, consists in the following extensions $12$, ${}^312$, $12^3$, $\cyc{123}$, $3123$, ${}^23123$, $3123^1$, ${}^23123^1$.

If were to glue only on $?12?$, not on all extensions thereof, we
would only generate the rule $T \to 231$, \ie\ the case where the bond
happens to belong to a triangle.  Clearly, this is a valid refinement
but is far from covering all cases.  Note also that the extension
$\phi_{1,2}:{} ?12? {}\to 12$ is not a minimal balanced extension as it
factors through either $\phi_{1,?}:{} ?12? {}\to 12?$ or $\phi_{?,2}:{}
?12? {}\to{} ?12$ which are both minimal.  This illustrates the
problem with prime extensions discussed earlier: the candidate
refinement $\{12?,?12\}$ is not valid precisely because it leads to an
ambiguous decomposition of the case of $\phi_{1,2}$ and will therefore
generate rates that are different for instances of $g_{12}$ with the
same balance vector $\Da \phi$.

On the other hand, the subrules
$3123$, ${}^23123$, $3123^1$, and ${}^23123^1$ all share the same balance vector. 
In fact, they can be summarized by a single prime extension $\phi_{?3,3?}:{}
?12? {}\to{} ?3123?$ which, together with $\phi_{T}:{} ?12? {}\to T$,
mutually exclusively and exhaustively refines
$\phi_{{}^312^3}:{} ?12? {}\to {}^312^3$.  Our growth policy is 
too local to see this optimisation, and has to refine $\phi_{?3,3?}$ into 
the above four explicit subcases. 

That said, the output of the growth policy could be either optimized by
hand or indeed subjected to the procedure of rule
{compression} which would automatically perform this optimization
for us.
A working Kappa model where the generators of this example have been
fully expanded according to the growth policy and then manually
compressed can be found in Appendix~\ref{app:triangles}.

\subsection{Rates}
To equip $\generators_{\shapes}$ with rates,
%
we suppose given a $\shapes$-indexed real-valued vector of \emph{energy costs} $\costs$, and 
a {rate map} $k:\generators_{\shapes}\to\mbb R_+$ such that, 
for all $g_\phi$ in $\generators_{\shapes}$:
\EQ{
\label{compat}
\log{k(g\st_{\phi\st})}-\log{k(g_\phi)}
&{=}&
\eps\cdot\Da\phi
}
with $\Da\phi$ in $\mbb Z^\shapes$, the balance vector of the refined rule 
$g_\phi$ with respect to $\shapes$,
a well-defined quantity by Th.~\ref{main}.

We write $\shapes(x)$ for the $\shapes$-indexed vector which maps $c$
to $|\mOR_C(c,x)|$, and define the \emph{energy} $E(x)$ of $x$ as
$\costs\cdot\shapes(x)$, \ie\ the sum over all $c \in \shapes$ of the
product of the number of instances of $c$ in $x$ and the energy cost
of $c$.
We also write $\LTS_\generators(x)$ for the finite (strongly) connected component of $x$ in $\LTS_\generators$,
and define a probability distribution (in Boltzmann format) $\pi_x$ on $\LTS_\generators(x)$ by:
\EQ{
\label{boltz}
\pi_{x}(y)&:=&
\dfrac
{\mathrm{e}^{-\eps\cdot \mcl P(y)}}
{\sum_{z\in \LTS_\generators(x)}\mathrm{e}^{-\eps\cdot \mcl P(z)}}
}
We can now prove our main theorem.


\begin{thm}
\label{conv}
Let $\generators$, $\shapes$, $\generators_{\shapes}$, $k$, and $\pi_x$ be defined as above; then (i)
$\LTS_{\generators_{\shapes}}$ and $\LTS_\generators$ are isomorphic as 
symmetric LTSs;
and (ii) 
for any mixture $x$,
the irreducible time-homogeneous continuous-time Markov chain $\LTS^k_{\generators_{\shapes}}$
has detailed balance for, and converges to, $\pi_x$
on $\LTS_{\generators_{\shapes}}(x)$. 
\end{thm}
\PR 
Both $\LTS_\generators$ and $\LTS_{\generators_{\shapes}}$ offer transitions from a mixture $x$:
the former are labelled by pairs $(g,\psi)$ with $g$ in $\generators$, $\psi$ in $\mOR_C(g_L,x)$;
the latter by pairs $(g_\phi,\ga)$ with $g_\phi$ the refinement of $g$ along a mature 
extension $\phi:g_L\to t$,
and $\ga$ in $\mOR_C(t,x)$.
Steps in the latter can be mapped to steps in the former
by transforming labels as follows: $(g_\phi,\ga) \mapsto (g,\ga\phi)$.
As $\generators_{\shapes}$ refines $\generators$ exhaustively (Th.~\ref{main}), this correspondence is a bijection, 
which establishes the first claim. 

(Pedantically, there is a full and faithful functor 
between the two corresponding free categories which is the identity on objects ---%
incidentally, this bijection is readily seen to respect the
symmetries on labels.)

Since we have multiple rules in $\LTS_{\generators_{\shapes}}$, each of which can be applied in several ways, there can be more than one transition from $x$ to the same $y$ ---each uniquely described by a $(g_\phi,\ga)$ label. 
Each such $(g_\phi,\ga)$ has an inverse, $(g\st_{\phi\st},\ga\st)$, 
where: 
$g\st$ is the rule inverse to $g$; 
$\phi\st$ corresponds to $\phi$ in the isomorphism between the categories of extensions 
of $x$ and $y$, with $\phi_\agents=\phi\st_\agents$; and 
$\ga\st$ is the embedding corresponding to $\ga$,
also with $\ga_\agents=\ga\st_\agents$.  
One can easily verify that $\phi\st$ is an epi, and
that $\phi\st$ is also mature. 
Hence $(g\st_{\phi\st},\ga\st)$ determines a valid transition
in $\LTS_{\generators_{\shapes}}$ which is inverse to $(g_\phi,\ga)$,
and we have a bijective correspondence
between transitions from $x$ to $y$ and those from $y$ to $x$.

Consider a pair $e$, $e\st$ of such corresponding events
due to $g_\phi$ and $g\st_{\phi\st}$; because 
$e$ is a transition from $x$ to $y$, and $\phi$
is $\shapes$-balanced (Th.~\ref{main}), we have 
$\shapes(y) = \shapes(x) + \Da\phi$, and hence
$\eps\cdot\Da\phi=\eps\cdot(\shapes(y)-\shapes(x))$;
so, by (\ref{compat}), the rates of $e$, $e\st$ are such that:
\AR{
k(e\st)\, 
\mathrm{e}^{-\eps\cdot\mcl P(y)} 
&{=}& 
k(e)\, 
\mathrm{e}^{-\eps\cdot\mcl P(x)}
}
and by summing this equation over all pairs, we obtain
detailed balance for the probability local to 
the component $\LTS_{\generators_{\shapes}}(x)=\LTS_{\generators_{\shapes}}(y)$,
defined above as $\pi_x=\pi_y$, since:
\AR{
q(y,x)\, 
\mathrm{e}^{-\eps\cdot\mcl P(y)} 
&{=}& 
q(x,y)\, 
\mathrm{e}^{-\eps\cdot\mcl P(x)} 
}
The convergence statement follows by standard results on finite CTMCs.
\RP


Note that the subset of the state space which is reachable from $x$ in $\LTS_{\generators}$, namely $\LTS_{\generators}(x)$, is finite; hence, 
the \emph{partition function} $Z(x):=\sum_{z\in\LTS_{\generators}(x)}\mathrm{e}^{-E(z)}$ which figures in the denominator of $\pi_x$
is also finite. In the presence of rules which increase the number of agents, the
components $\LTS_{\generators}(x)$ can be infinite and $Z(x)$ may diverge. For (mass action stochastic) Petri nets, convergence is guaranteed if detailed balance holds, but it is not true in general for Kappa~\cite{et2,et1}.

Another point worth making is that the result holds
symbolically ---regardless of the energy costs $\costs$.
So $\costs$ can be seen as a set of parameters,
an ideal support for machine learning techniques
if one were contemplating fitting a model to data.

\subsection{A linear kinetic model}\label{kin}

We now ask: what of the \emph{actual} rates of $\LTS^k_{\generators_{\shapes}}$? 
Among all possible choices which accord with (\ref{compat}), we 
delineate a tractable subset whose size grows quadratically in $|\shapes|$. This is a 
useful log-linear heuristic, which is common in machine learning, but 
has no particular claim to validity.


Suppose we have, for every generating rule $g$ in $\mcl G$, 
a constant $c_g\in\mbb R$, 
and a matrix $A_g$ of dimension $|{\shapes}|\times|{\shapes}|$.
Subject to the constraints that 
$c_{g\st}=c_g$, and $A_{g\st}+A_g=I$, 
we can define a log-affine rate map 
which satisfies (\ref{compat}) by:
\EQ{
\label{KM1}
\log(k(g_\phi))
&:=& 
c_g - A_g(\eps)\cdot\Da\phi
}
%
%
%
%
The kinetic model 
expressed in (\ref{KM1}) requires of the order of $|\shapes|^2\times|\mcl G|$ parameters.
In practice, one needs even fewer parameters, as only those energy patterns that 
are relevant to a given $g$, \ie\ have non-zero balance for at least one rule in $\Ga^\shapes(g)$, 
need to be considered when building $A_g$. 
Typically, for larger models, this will be a far smaller number than $|\shapes|$.
This relative parsimony is compounded by the fact that the number of \emph{independent} parameters will be often lower, because the $\Da\phi$ family often has low rank. It is to be compared with the total number of choices which is far greater as it is of the order of the number of refinements, that is to say $\sum_{g\in\mcl G}|\Ga^\shapes(g)|$.

If we set $c_{g\st}=c_g=0$, $A_{g\st}=0$,
$A_g=I$, we get: $k(g_\phi)=\mathrm{e}^{-\eps\cdot\Da\phi}$,
$k(g\st_{\phi\st})=1$. As $\eps\cdot\Da\phi$ is the difference of energy between the target and source in any application $g_\phi$, 
this choice amounts to being exponentially reluctant to 
climb up the energy gradient. This is a 
continuous-time version of the celebrated Metropolis algorithm~\cite{metro}.


Another particular case, completely symmetric and which can be a reasonable choice to begin with, 
is obtained for 
$A_{g\st}=A_g=I/2$:
\AR{
\label{mid}
k_g(\phi) &=& C_g\,e^{-\eps\cdot\Da\phi/2}\\
k_{g\st}(\phi\st) &=&C_g\,e^{\eps\cdot\Da\phi/2}
}
with $C_g=e^{c_g}$. As $\Da\phi\st=-\Da\phi$, this is indeed a symmetric definition.

%



Finally, it is fun to draw a comparison between the ascription given in  (\ref{KM1}) and 
the Arrhenius rate law.
This law posits a dependency of the rate constant $k$ of a reaction of the form 
$\log k = c  - E_a/kT$, where $c$ is a constant (defining the basic time scale
of the reaction),
$E_a$ is the so-called \emph{activation energy} of the reaction
and $T$ is the temperature. In our case, we are not concerned with the effect of
$T$ on the (logarithm of the) rate but with the effect of consuming and
producing various energy patterns in $\shapes$ 
at the locus of the instance of the generator rule $g$. In this view of things,
(\ref{KM1})
posits that the `activation energy' of $\phi$ depends linearly on the cost of
the various patterns and the balance of $\phi$.

\subsection{Energy functions do not need to be linear}
We now return to a key assumption made in the preceding section and
consider a more general situation where the energy function $E$ is no
longer asked to be linear. For reasons to become clear shortly, we
still assume the much weaker property 
that $E$ can be factored as $v \circ \shapes(\_)$
for some finite set of patterns $\shapes$.
\EQ{\label{factor}
\xymatrix@R=15pt{
\tbf{rSGe}_C\ar[r]^{\shapes(\_)}&
\tbf{mSet}(\shapes)\ar[r]^>>>>>{v}&
\mbb R
}
}
Note that if we see multi-sets as equipped with the usual point-wise
partial order, $\shapes(\_)$ is evidently functorial.

As an example, consider a divalent agent with contact graph
$X(a^1,b^1)$ ---where the shared superscript symbolizes an edge between
sites $a$ and $b$--- with which one can form only chains and cycles,
and a single generator $g$ which can create/delete the unique edge
type.  Write $c_3$ for a cycle of length $3$ (a triangle), and $t_2$
for an open chain of length $2$.  We define the quadratic energy
function $E(x) = |\mOR_C(c_3,x)|^2$, \ie\ $v(n)=n^2$.  Applying $g$ forward
to a chain $t_2$ in a site graph of the form $x=t_2+x'$ 
will create a new copy of $c_3$, and give the following energy balance:
\AR{
\Da E &=& 
(|\mOR_C(c_3,x)|+1)^2-(|\mOR_C(c_3,x)|)^2
&=&2|\mOR_C(c_3,x)| + 1
}
Therefore, detailed balance forces the log-ratio $\hat K_g$ of the
backward/forward rates assigned to an edge creation to {depend} on
$x$.  This is unlike the case of linear potentials where this ratio is
independent of $x$.
However, this extension $g_\phi$ of $g$ (where it is applied to a
chain $t_2$) is balanced with respect to $\shapes$.  This means, as we
have seen, that the stoichiometric $\shapes$-vector $\Da\phi$
associated to $g_\phi$ ---where each component $\Da\phi(c)$ is defined
as the difference of $|\mOR_C(c,y)|-|\mOR_C(c,x)|$ for a
$g$-transition from $x$ to $y$--- is the same for all $x,y$.
In the example $\Da\phi$ has only one component, which is constant
$\Da\phi(c_3)=1$ because the binding of the two \emph{free} extremes
of an open chain $t_2$ can only produce one triangle, regardless of
the context in which the refined rule $g_\phi$ is applied.

In general, one can visualize the situation as follows.
\AR{
\xymatrix@C=40pt@R=30pt{
x\ar[r]^{g}\ar[d]_{\shapes}&y\ar[d]^{\shapes}\\
{\shapes(x)}\ar[r]^{+\Da\phi}\ar[d]_v&{\shapes(y)}\ar[d]^v\\
\mbb R\ar[r]^{+\Da E}&\mbb R
}
}
and detailed balance amounts to asking for $\hat K_g = v(\shapes(x) +
\Da\phi(x)) - v(\shapes(x))$.  If $v$ happens to be linear then this
is the usual condition $\hat K_g = v(\Da\phi(x))$. If $v$ is not
linear, the constraint does not seem very helpful as \emph{a priori}
one has to know $x$ to compute the right hand side. But by the
assumption (\ref{factor}) opening the paragraph, $\Da\phi$ factors
through $\shapes$, hence:
\AR{
\hat K_g &=& v(\shapes(x)+\sig_g(\shapes(x)))-v(\shapes(x))
&=:& \psi_g\circ\shapes(x)
}
where the second equation defining $\psi_g$ uniquely as a real-valued
function from $\shapes$-multisets.  With this rewriting, it is plain
that the constraint depends not on full knowledge of $x$, but only on
$\shapes(x)$.  Equivalently, we see that $\hat K_g$ factors through
$\shapes(\_)$ just like $E$.  In the example, $\hat K_g=2|\mOR_C(c_3,x)|+1$,
and $\psi_g(n)=2n+1$.

This is good enough to define rates for a balanced $g_\phi$.
For example, by analogy with the earlier linear kinetic model, we can
choose log-rates (seen as real-valued multi-set functions) as follows:
\EQ{
\label{kinmod}
\hat k_g = \al_g - \ba_g \psi_g
}
with $\al_g$, $\ba_g$ real-valued functions on $\shapes$-multisets
such that $\al_{g\st}=\al_g$ and $\ba_{g\st}+\ba_g=1$.
This assignment solves the above constraint as, clearly,
$\psi_{g\st}+\psi_g=0$.

From the \emph{simulation} point of view, this added generality
requires two things: (i) that rates can be made to depend explicitly
on observables; (ii) that the internal state of the simulation be
extended to incorporate $\shapes(x)$.  Both possibilities are
already generically available in the current version of the main Kappa
simulator KaSim~\cite{KaSimManual2014}.
A slight modification of the engine (not implemented) could obtain
direct updates to $\shapes(x)$ as, by assumption, applying $g_\phi$
leads to a constant $+\Da\phi$ update; and the same holds for
propagating these updates to the rates of the rules which depend on
them, \eg\ as in (\ref{kinmod}).  Thus, the complexity properties of
the simulation algorithm are preserved~\cite{smbp:07}.

\section{Allosteric ring}\label{alloring}

We now put our methodology to use on a realistic example
of a bacterial flagellar engine.
In this section, we will use the traditional syntax of Kappa to denote
site graphs: subscripts for states and shared superscripts for edges
between sites, \eg\ $A(x_0^1),B(y^1)$.  Unlike the mathematical
definitions of \S\ref{rewriting}, agent and site types are indicated
as explicit labels.
%
Again, we use \href{https://github.com/Kappa-Dev/KaSim}{KaSim} (the standard
Kappa engine) for the simulation shown below.

\subsection{The model}


The engine can rotate clockwise or anti-clockwise at high angular velocities, and this will decide whether the bacterium tumbles or swims forward.
%
One can build a simple model of the switch between the two modes~\cite{teuta}.
The engine is seen as a ring of $n$ identical components, $P$, with two possible conformations, $0$ and $1$.
(In reality, each of the $n=34$ component protomers is itself a tiny complex made of different subcomponents, but the model ignores this.)   
A ring homogeneously in state $0$ ($1$) rotates (anti-) clockwise and induces tumbling (straight motion). Importantly, neighbouring $P$s on the ring prefer to have matching conformations. 
States of the ring with many mismatches thus incur high penalties. 
In the absence of any \emph{Y} molecule binding a $P$, its favoured conformation is $0$; conversely, in the presence of a \emph{Y}, \emph{P} favours $1$. ($Y$ stands for a small diffusible protein named \emph{CheY}.) To bind, \emph{Y} has to be activated by an external signal. Hence the switch can be triggered by a sudden activation of \emph{Y} which then binds the ring and induces a change of regime. The sharper the transition between the two regimes the better.

As each of the $P$s can be in four states, the ring has on the order of $10^{20}$ non-isomorphic configurations which precludes any reaction-based (\eg\ Petri nets) approach to the dynamics where each global state is considered as  one chemical species. At this stage, we could apply the rule-based approach, 
or, better, we can obtain the rules \emph{indirectly} by applying the
methodology of \S\ref{rulegen}.  This is what we now do informally.




First, we define our contact graph with two agent types: 
$P(x,y,f_{0,1},s)$ with domains $x$, $y$ to form the ring, $s$ to bind its signal $Y$, and $f$ a placeholder for $P$'s conformation; $\emph{Y}(s_{u,p})$ with two internal states to represent activity.

\begin{wrapfigure}[7]{r}{0.42\textwidth}
\begin{minipage}{0.42\textwidth}
\AR{
\emph{Motif}
&\qquad
\emph{Cost}
\\{}\\
P(f_i,x^1),P(y^1,f_j)
&\qquad
\eps^{PP}_{ij} 
\\
P(f_i)
&\qquad\eps^{P}_{i}
\\
P(f_i,s^1),Y(s^1)
&\qquad \eps^{PY}_{i}
}
\end{minipage}
\end{wrapfigure}

Second, we capture the informal statements in the discussion above by
defining the energy patterns and associated costs.  Note that the
various motifs overlap.  Following \S\ref{rulegen}, we associate to
each ring configuration $x$ the occurrence vector $\shapes(x)$ and
total energy $\costs \cdot \shapes(x)$.
%
For example, a ring of size $n$ uniformly in state $0$ and with no bound $Y$s has total energy $n(\eps^{PP}_{00}+\eps^{P}_{0})$. This, in turn, defines the equilibrium distribution of the ring, namely $x$ has probability proportional to $\exp(-\eps\cdot \shapes(x))$. 
(The convention is that the lower the energy, the likelier the state.)

\begin{wrapfigure}[6]{r}{0.36\textwidth}
\vspace{-1.5ex}
\begin{minipage}{0.36\textwidth}
\EQ{
\label{ePP}
\eps^{PP}_{00},\eps^{PP}_{11} &<&\eps^{PP}_{10},\eps^{PP}_{01}
\\
\label{eP}
\eps^{P}_{0} &{<}& \eps^{P}_{1} 
\\
\label{ePY}
\eps^{PY}_{0} &{>}& \eps^{PY}_{1} 
}
\end{minipage}
\end{wrapfigure}
In order to complete our energy landscape, 
we need to pick energy costs which reward or penalize local configurations 
as discussed above:
the role of (\ref{ePP}) is to align the internal states of neighbours on the ring
---an Ising penalty term for mismatching neighbours which will ``spread conformation''; 
(\ref{eP}) makes $0$ the favoured state, while (\ref{ePY}), which 
kicks in only in the presence of \emph{Y}, makes $1$ the favoured state.



The next step is to create the dynamics. The naive rule $b$ for \emph{PY} binding:
\AR{
b := P(s),Y(s_p) \lrar P(s^1),Y(s_p^1) 
}
has a $\Da E$ which is ambiguous as it will be either $\eps^{PY}_0$ or $\eps^{PY}_1$ depending on its instances;
hence, we have no hope of assigning rates to this rule that satisfy detailed balance ---unless $\eps^{PY}_0=\eps^{PY}_1$, which contradicts
(\ref{ePY}). To get a definite balance, one needs to refine this rule:
\AR{
b_0 := P(f_0,s),Y(s_p) \lrar P(f_0,s^1),Y(s_p^1)\\
b_1 := P(f_1,s),Y(s_p) \lrar P(f_1,s^1),Y(s_p^1)
}
Now each rule $b_i$ specifies enough of
the context in which it applies to have a definite energy balance $\eps^{PY}_i$. Following the same intuition of revealing (just) enough context, 
we obtain a balanced rule set for conformational changes:
\AR{
f_{ij} 
&{:=}& 
P(f_i,y^1),P(x^1,f_0,y^2,s),P(x^2,f_j) \lrar
P(f_i,y^1),P(x^1,f_1,y^2,s),P(x^2,f_j)
\\
f'_{ij} 
&{:=}& 
P(f_i,y^1),P(x^1,f_0,y^2,s^\_),P(x^2,f_j) \lrar
P(f_i,y^1),P(x^1,f_1,y^2,s^\_),P(x^2,f_j)
}
The first (second) group of rules represents the changes in the absence (presence) of a $Y$ bound to the middle $P$ undergoing a change of conformation. (The fact that $P$'s site $s$  is bound is indicated by the underscore exponent.) 

These $f$-rules have respective balance:
\AR{
\eps^{PP}_{i1}+\eps^{PP}_{1j}-\eps^{PP}_{i0}-\eps^{PP}_{0j}
+\eps^{P}_{1}-\eps^{P}_{0}
\\
\eps^{PP}_{i1}+\eps^{PP}_{1j}-\eps^{PP}_{i0}-\eps^{PP}_{0j}
+\eps^{P}_{1}-\eps^{P}_{0}
+\eps^{PY}_{1}-\eps^{PY}_{0}
}
As we have ten reversible rules, and only eight energy patterns, there must
be linear dependencies between the various balances. Indeed, in this case, it is
easy to see that the family of vector balances has rank six. Thermodynamic
consistency induces relationships between rates; a well-established fact 
in the case of reaction networks (\eg\ see Ref.~\cite{et2}).

\begin{figure}[t!]
   \centering
   \includegraphics[width=350pt]{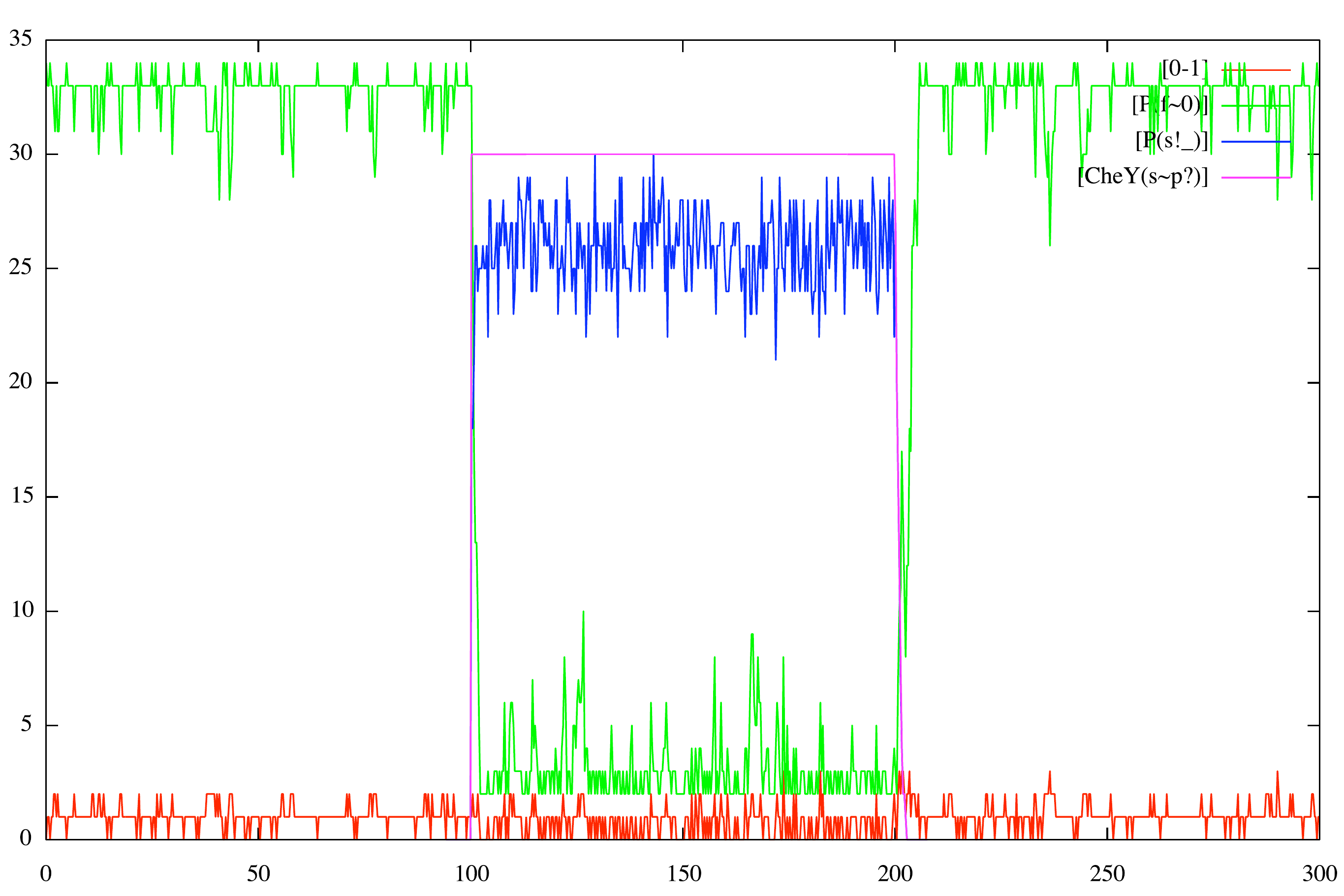} 
   \caption{The simulation steps up the amount of active $Y$ at $t=100$, and down again at $t=200$; 
this sends the entire ring into an homogeneously $1$ conformation, and back to $0$. The number of mismatches (lowest curve) stays low, even during transitions.}
   \label{PY}
\end{figure}

With the rules in place, 
the final step is to choose rates
which satisfy detailed balance. This guarantees
that the obtained rule set converges to the equilibrium specified by the choice of 
the energy cost vector. 
Convergence will happen whatever $\eps$ is, \ie\ symbolically. If, in addition, 
$\eps$ follows (\ref{ePP}--\ref{ePY}), one can see in Fig.~\ref{PY} that the ring
(i) undergoes sharp transitions when active \emph{Y} is stepped up and down again;
and (ii) has at all times very few mismatches. 





%

\subsection{How to generate the rules}
Our set of  balanced rules 
was based on two generators, $b$ for binding, $f$ for flipping:
\AR{
b := P(s),Y(s_p) \lrar P(s^1),Y(s_p^1)
\\
f := P(f_0)\lrar P(f_1)
}
Note that there is a design choice here. In effect, we are saying that we are not interested
in forming/breaking the bonds between the $P$s in the ring. 
If we wanted to incorporate also the ring assembly in the model, we would have to add $P(x),P(y) \lrar P(x^1),P(y^1)$ among our 
generator set $\generators$. This would generate many more refined rules, as we will see.
Recall that our patterns fall in three subgroups: 
$P(f_i,x^1),P(y^1,f_j)$;
$P(f_i)$;
and $P(f_i,s^1),Y(s^1)$.

Consider the extensions of $b$:
clearly only the last pattern can glue relevantly on it; the corresponding
(unique) site request is for $P$ to reveal $f$ and its internal state. This gives the
first two rules $b_0$, $b_1$. 

Consider now the more interesting extensions of $f$:
the second pattern type glues relevantly but does not generate any site request;
the third one asks $P$ to reveal its site $s$, resulting
in two possible extensions ($s^\_$ means that $s$ is bound):
\AR{
P(f_0,s)\lrar P(f_1,s)
\\
P(f_0,s^\_)\lrar P(f_1,s^\_)
}
These extensions are \emph{not} mature yet, as one can glue relevantly  patterns 
of the first type on both sides of $P$, inducing a further request for 
revealing $P$'s sites $x$ and $y$.

If we are in
the component of an initial state where $P$s are arranged in a ring,
then we know that the neighbours on both sides exist and are $P$s; this
gives the final refinement of the above into the
rules $f_{ij}$, $f'_{ij}$ described earlier. If, on the other hand, we do
not know that, we also have to add several rules where one or both of $x$,
$y$ are free, corresponding to open $P$-chains. This demonstrates the sensitivity
of the obtained rule set to the initial choice of generators.

Hence, the rules we generated by hand are indeed the ones we would generate using our general refinement strategy of \S\ref{rulegen}.
%

\begin{figure}[t]
   \centering
   \includegraphics[width=350pt]{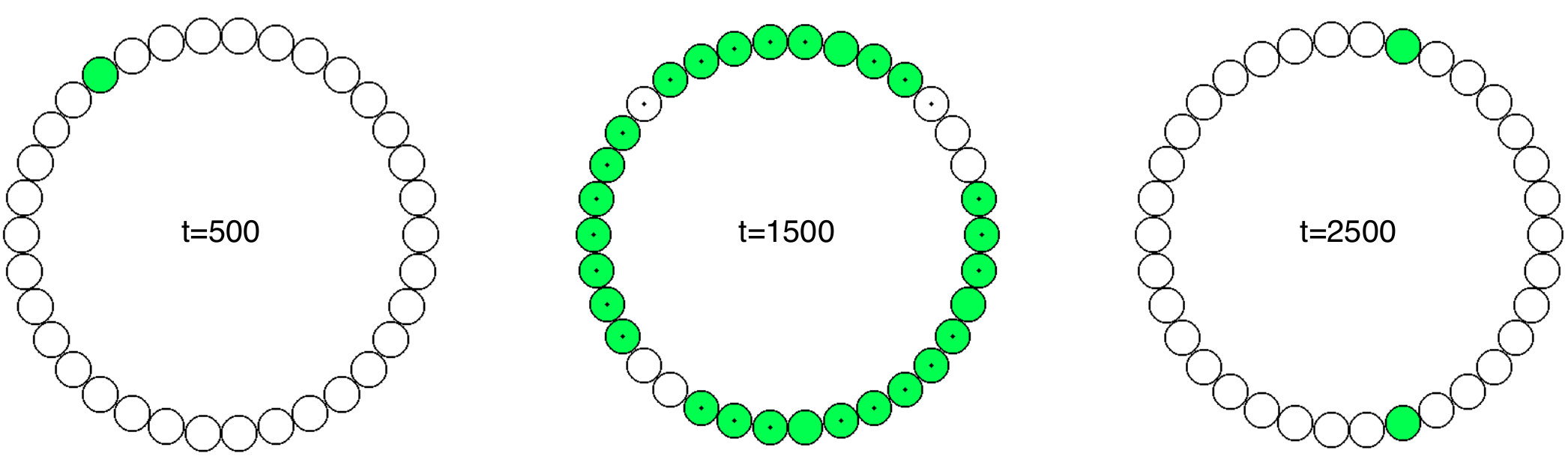} 
   \caption{Snapshots of the ring configuration are taken at time $500$, $1500$, and $2500$. Solid (green) circles indicate conformation $1$, hollow ones conformation $0$; a dot in the centre indicates a bound (hence active) $Y$. At times $500$, $2500$, no $Y$ is bound (because they are all inactive) and the ring is globally in state $0$, up to tiny fluctuations; at time $1500$, it is globally in state $1$ as a consequence of the binding of $Y$s.}
   \label{snap}
\end{figure}

We can visualize the obtained simulations  by extracting snapshots before, after and during the injection of active $Y$s, as in Fig.~\ref{snap}. Again we see few mismatches in both regime because of the Ising interaction expressed by the $\eps^{PP}$ energy costs.
The choice of rates made in Ref.~\cite{teuta} for the $f$-generator 
is the symmetric version of (\ref{KM1}).

\section{Conclusion}

We have presented a new `energy-oriented' methodology for the development of site graph rewriting models based on a set $\shapes$ of energy patterns; these patterns use a graphical syntax which allows us to specify the energy landscape. Rewrite rules are implicitly defined by 
$\shapes$ and generator rules $\generators$. The resulting rule set $\generators_\shapes$ is guaranteed to be thermodynamically correct and to converge to the probability distribution described by the energy landscape, given suitable rates. The construction is entirely parametric in the energy costs $\eps$, and modular in $\generators$. This means that in a modelling context, one can sweep over various values for $\eps$ without having to rebuild the model, and compositionally add new rule components to $\generators$. Both features are clearly useful. 
%
%
%
We expect our construction to provide a broad and uniform language to describe and analyse models of interacting biomolecules and similar systems of a quantitative fine-grained and distributed nature. 


There are no specific conditions bearing on this construction other than that energy patterns should be local. It would be interesting to investigate whether suitable constraints on patterns and generator rules can obtain optimized generated rule sets. Another interesting
extension would be to deal with non-local forms of energies expressing
long-range interactions, where the metric is read off the graph itself. In practice, there will be many more rules generated,
and beyond the descriptive aspects, simulations will need new ideas to be feasible. A ray of hope comes from the log-affine kinetic model
presented in \S3, as rules can be partitioned by energy balances for faster selection. 


Finally, as said in the introduction, there is a growing body of literature which turns a theoretical eye to site graph rewriting~\cite{jonandtobias,dixon,heckel,kappadpo}, and it is tempting to ask whether our derivation can be replayed in more abstract settings; in particular, it would be very interesting to investigate its integration with the abstract framework for rule-based modelling developed in \cite{lynch}. 

%

\paragraph{Acknowledgements}
We would like to thank Peter Swain, Andrea Weisse, Julien Ollivier,
Nicolas Oury, Finlo Boyde and Eric Deeds for many useful conversations
concerning the subject of this paper.
VD acknowledges the support of the ERC 320823 RULE project.

This work was sponsored by
the Defense Advanced Research Projects Agency (DARPA) and
the U. S. Army Research Office under grant number W911NF-14-1-0367.
The views, opinions, and/or findings contained in this report
are those of the authors and should not be interpreted as representing
the official views or policies, either expressed or implied,
of the Defense Advanced Research Projects Agency or the Department of Defense.

\small
\bibliographystyle{plain}
\bibliography{papers,vd}


\normalsize

\newpage

\appendix

\section{Triangles model}\label{app:triangles}

A working model definition for our first example (introduced in
\S\ref{triangles}) is spelled out in full here.  The syntax used is
for \href{https://github.com/Kappa-Dev/KaSim}{KaSim} version 4.
As in the example of \S4, we use the symmetric linear kinetic model of \S\ref{kin}.

\subsection{Agents, Parameters and Initial State}
We start by declaring agents together with their sites.
Then we declare the energy costs for $T = \cyc{123}$ (`t'), $?12?$
(`ab'), $?23?$ (`bc'), $?31?$ (`ca') as parameters.
Subsequently the initial state is defined as containing $1000$ agents
of each type.

\begin{verbatim}
# Agent signatures
%agent: A(l,r)
%agent: B(l,r)
%agent: C(l,r)

# Energy costs
%var: 't' -10
%var: 'ab' 0
%var: 'bc' 0
%var: 'ca' 0

# Initial state
%init: 1000 (A(), B(), C())
\end{verbatim}

\subsection{Observables}
Next we declare the graphs we would like to keep track of, \ie\ those
graphs we would like to generate trajectories for plotting.
Here we are interested in the fraction of agents that are used to
assemble triangles.

\begin{verbatim}
%obs: 'T'  |A(l!1, r!2), B(l!2, r!3), C(l!3, r!1)|
\end{verbatim}

\subsection{Rules}
Finally, we write our set of rules.
These have been grouped according to the qualitative generator rule
they refine.  Note that these rules have been manually compressed
as explained in \S\ref{triangles-detail}.

\begin{verbatim}
# A(r), B(l) -> A(r!1), B(l!1) refines into:
A(l,r), B(l,r) -> A(l,r!1), B(l!1,r) @ [exp] (-1/2 * 'ab')
A(l!r.C,r), B(l,r) -> A(l!r.C,r!1), B(l!1,r) @ [exp] (-1/2 * 'ab')
A(l,r), B(l,r!l.C) -> A(l,r!1), B(l!1,r!l.C) @ [exp] (-1/2 * 'ab')
A(l!1,r  ), B(l  ,r!3), C(l!3,r!1) -> \
A(l!1,r!2), B(l!2,r!3), C(l!3,r!1) @ [exp] (-1/2 * ('ab' + 't'))
C(r!1), A(l!1,r  ), B(l  ,r!3), C(l!3) -> \
C(r!1), A(l!1,r!2), B(l!2,r!3), C(l!3) @ [exp] (-1/2 * 'ab')

# A(r!1), B(l!1) -> A(r), B(l) refines into:
A(l,r!1), B(l!1,r) -> A(l,r), B(l,r) @ [exp] -(-1/2 * 'ab')
A(l!r.C,r!1), B(l!1,r) -> A(l!r.C,r), B(l,r) @ [exp] -(-1/2 * 'ab')
A(l,r!1), B(l!1,r!l.C) -> A(l,r), B(l,r!l.C) @ [exp] -(-1/2 * 'ab')
A(l!1,r!2), B(l!2,r!3), C(l!3,r!1) -> \
A(l!1,r  ), B(l  ,r!3), C(l!3,r!1) @ [exp] -(-1/2 * ('ab' + 't'))
C(r!1), A(l!1,r!2), B(l!2,r!3), C(l!3) -> \
C(r!1), A(l!1,r  ), B(l  ,r!3), C(l!3) @ [exp] -(-1/2 * 'ab')

# B(r), C(l) -> B(r!1), C(l!1) refines into:
B(l,r), C(l,r) -> B(l,r!1), C(l!1,r) @ [exp] (-1/2 * 'bc')
B(l!r.A,r), C(l,r) -> B(l!r.A,r!1), C(l!1,r) @ [exp] (-1/2 * 'bc')
B(l,r), C(l,r!l.A) -> B(l,r!1), C(l!1,r!l.A) @ [exp] (-1/2 * 'bc')
B(l!1,r  ), C(l  ,r!3), A(l!3,r!1) -> \
B(l!1,r!2), C(l!2,r!3), A(l!3,r!1) @ [exp] (-1/2 * ('bc' + 't'))
A(r!1), B(l!1,r  ), C(l  ,r!3), A(l!3) -> \
A(r!1), B(l!1,r!2), C(l!2,r!3), A(l!3) @ [exp] (-1/2 * 'bc')

# B(r!1), C(l!1) -> B(r), C(l) refines into:
B(l,r!1), C(l!1,r) -> B(l,r), C(l,r) @ [exp] -(-1/2 * 'bc')
B(l!r.A,r!1), C(l!1,r) -> B(l!r.A,r), C(l,r) @ [exp] -(-1/2 * 'bc')
B(l,r!1), C(l!1,r!l.A) -> B(l,r), C(l,r!l.A) @ [exp] -(-1/2 * 'bc')
B(l!1,r!2), C(l!2,r!3), A(l!3,r!1) -> \
B(l!1,r  ), C(l  ,r!3), A(l!3,r!1) @ [exp] -(-1/2 * ('bc' + 't'))
A(r!1), B(l!1,r!2), C(l!2,r!3), A(l!3) -> \
A(r!1), B(l!1,r  ), C(l  ,r!3), A(l!3) @ [exp] -(-1/2 * 'bc')

# C(r), A(l) -> C(r!1), A(l!1) refines into:
C(l,r), A(l,r) -> C(l,r!1), A(l!1,r) @ [exp] (-1/2 * 'ca')
C(l!r.B,r), A(l,r) -> C(l!r.B,r!1), A(l!1,r) @ [exp] (-1/2 * 'ca')
C(l,r), A(l,r!l.B) -> C(l,r!1), A(l!1,r!l.B) @ [exp] (-1/2 * 'ca')
C(l!1,r  ), A(l  ,r!3), B(l!3,r!1) -> \
C(l!1,r!2), A(l!2,r!3), B(l!3,r!1) @ [exp] (-1/2 * ('ca' + 't'))
B(r!1), C(l!1,r  ), A(l  ,r!3), B(l!3) -> \
B(r!1), C(l!1,r!2), A(l!2,r!3), B(l!3) @ [exp] (-1/2 * 'ca')

# C(r!1), A(l!1) -> C(r), A(l) refines into:
C(l,r!1), A(l!1,r) -> C(l,r), A(l,r) @ [exp] -(-1/2 * 'ca')
C(l!r.B,r!1), A(l!1,r) -> C(l!r.B,r), A(l,r) @ [exp] -(-1/2 * 'ca')
C(l,r!1), A(l!1,r!l.B) -> C(l,r), A(l,r!l.B) @ [exp] -(-1/2 * 'ca')
C(l!1,r!2), A(l!2,r!3), B(l!3,r!1) -> \
C(l!1,r  ), A(l  ,r!3), B(l!3,r!1) @ [exp] -(-1/2 * ('ca' + 't'))
B(r!1), C(l!1,r!2), A(l!2,r!3), B(l!3) -> \
B(r!1), C(l!1,r  ), A(l  ,r!3), B(l!3) @ [exp] -(-1/2 * 'ca')
\end{verbatim}

\subsection{Results}
\begin{figure}
  \begin{center}
    \includegraphics[width=.9\textwidth]{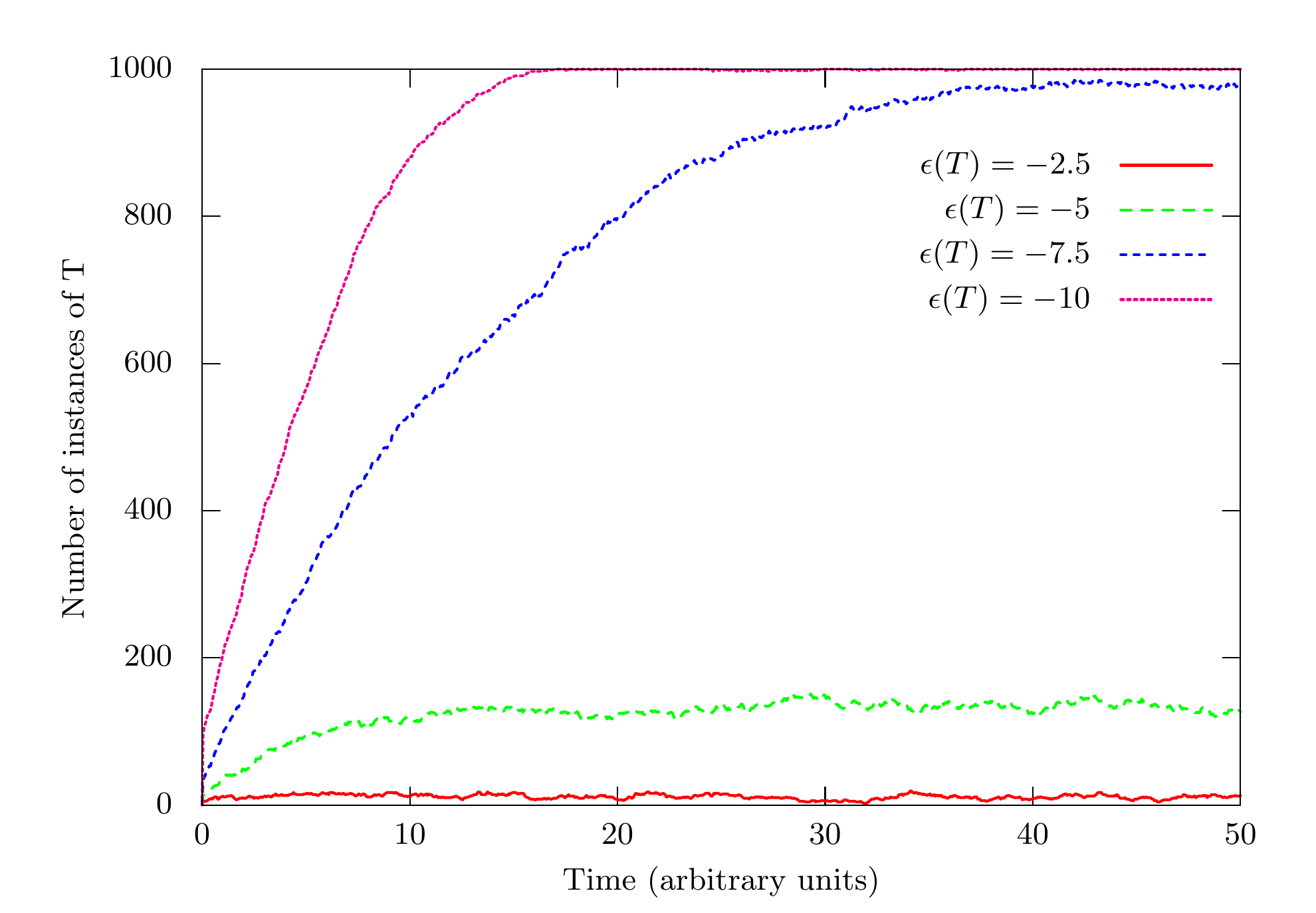}
  \end{center}
  \caption{Trajectories for $T = \cyc{123}$ when $\costs(T)$ varies
    (this value is set by changing the value of parameter `t' in the
     KaSim file).}
  \label{triangles-sim}
\end{figure}

We run this model using \href{https://github.com/Kappa-Dev/KaSim}{KaSim}
(version 4) to obtain trajectories for the number of triangles during
the simulation.  From these numbers we can estimate what is the
expected number of triangles at equilibrium.
In Fig.~\ref{triangles-sim} the results of 4 different runs with
different energy costs for a unit triangle are displayed.
When $\costs(T) = -10$, almost all agents are used to build triangles.
Instead, when $\costs(T) = -2.5$ only less than 5\% is used.
Interestingly, in both cases the set of states that minimize the
energy function is the same, namely those states that maximize the
amount of triangles.
So then why is it that in the latter case there are so few triangles?
The reason is entropic: although the probability of being in a state
with few triangles is small, there are many such states and together
they outweigh the probability of being in the few states were the
energy is minimized.
By further decreasing the energy of those few states we compensate
for this mass effect, until at $\costs(T) = -10$, order wins, and 
the effect is not noticeable anymore.

\end{document}